%% file: main.tex
\documentclass[12pt,twocolumn]{aastex62}

\usepackage{hyperref}
\usepackage{amsmath,mathtools,mathrsfs,amssymb}
\usepackage{graphicx}
\usepackage{bm}

\graphicspath{{./}{figures/}}

\revised{March 1, 2021}
\accepted{March 12, 2021}
\submitjournal{ApJ}

\shorttitle{Fast magnetic reconnection structures in Poynting-flux dominated jets}
\shortauthors{Kadowaki et al.}
\watermark{DRAFT}

\usepackage{multirow}

\defcitealias{medina_torrejon_etal2020}{MGK21}

\begin{document}

\title{Fast magnetic reconnection structures in Poynting-flux dominated jets}

\correspondingauthor{Luis H.S. Kadowaki}
\email{luis.kadowaki@iag.usp.br}

\author[0000-0002-6908-5634]{Luis H.S. Kadowaki}
\altaffiliation{FAPESP Fellow}
\affiliation{Universidade de S\~{a}o Paulo, Instituto de Astronomia, Geof\'{i}sica e Ci\^{e}ncias Atmosf\'{e}ricas, Departamento de Astronomia, 1226 Mat\~{a}o Street, S\~{a}o Paulo, 05508-090, Brasil}
             
\author[0000-0001-8058-4752]{Elisabete M. de Gouveia Dal Pino}
\affiliation{Universidade de S\~{a}o Paulo, Instituto de Astronomia, Geof\'{i}sica e Ci\^{e}ncias Atmosf\'{e}ricas, Departamento de Astronomia, 1226 Mat\~{a}o Street, S\~{a}o Paulo, 05508-090, Brasil}

\author{Tania E. Medina-Torrej\'{o}n}
\affiliation{Universidade de S\~{a}o Paulo, Instituto de F\'{i}sica, S\~{a}o Paulo, Brasil}

\author[0000-0002-8131-6730]{Yosuke Mizuno}
\affiliation{Tsung-Dao Lee Institute and School of Physics and Astronomy, Shanghai Jiao Tong University, Shanghai, 200240, People’s Republic of China}
\affiliation{Institut f\"{u}r Theoretische Physik, Goethe Universit\"{a}t, 60438 Frankfurt am Main, Germany}

\author{Pankaj Kushwaha}
\altaffiliation{Aryabhatta Postdoctoral Fellow}
\affiliation{Aryabhatta Research Institute of Observational Sciences (ARIES), Manora Peak, Nainital 263001, India.}



\begin{abstract}

The ubiquitous relativistic jet phenomena associated with black holes play a major role in high and very-high-energy (VHE) astrophysics. In particular, observations have demonstrated that blazars show  VHE emission with time-variability from days to minutes (in the Gev and TeV bands), implying very compact emission regions. The real mechanism able to explain the particle acceleration process responsible for this emission is still debated, but magnetic reconnection has been lately discussed as a strong potential candidate. In this work, we present the results of three-dimensional special relativistic magnetohydrodynamic simulations of the development of reconnection events driven by turbulence induced by current-driven kink instability along a relativistic jet. We have performed a systematic identification of all reconnection regions in the system, characterizing their local magnetic field topology and quantifying the reconnection rates. We obtained average rates of $0.051\pm0.026$ (in units of the Alfv\'{e}n speed) which are comparable to the predictions of the theory of turbulence-induced fast reconnection. Detailed statistical analysis also demonstrated that the fast reconnection events follow a log-normal distribution, which is a signature of its turbulent origin. To probe the robustness of our method, we have applied our results to the blazar Mrk 421. Building a synthetic light curve from the integrated magnetic reconnection power, we evaluated the time-variability from a power spectral density analysis, obtaining a good agreement with the observations in the GeV band. This suggests that turbulent fast magnetic reconnection can be a possible process behind the high-energy emission variability phenomena observed in blazars.

\end{abstract}



\section{Introduction} \label{sec:intro}

Relativistic jets are observed in different classes of astrophysical sources, from mildly- or highly-relativistic in microquasars (BH XRBs) and active galactic nuclei (AGNs), to ultra-relativistic ones in gamma-ray bursts (GRBs). The observation of polarized non-thermal radiation (from radio to gamma-rays) is suggestive of strong magnetization in these jets, particularly near the nuclear region \citep[e.g.,][]{laurent_etal_11, doeleman_etal_12, marti-vidal_etal_15}. There are even observational hints of helical magnetic field structures in these regions \citep[as observed, e.g., in M87,][]{harris_etal_03}.

The most accepted models to explain the origin of jets combine magnetic processes with rotation. In the seminal work of \citet{blandford1977}, jets can be powered by the black hole spin transferred to the surrounding magnetic flow. Alternatively, jets can be driven by magneto-centrifugal acceleration in helical magnetic fields arising from the accretion disk \citep{blandford1982}. In both cases the prediction is that jets should be born as magnetized flows. 

Observations also indicate that at distances large enough from the source, 
corresponding to scales of  several orders of magnitude of  the Schwarzschild radius, these jets should be already kinetically dominated \citep[see, e.g.,][]{nakamura_asada_13, zamaninasab_etal_14, christie_etal_19, giannios_uzdensky2019, zhang_giannios2021},
so that there should be some efficient conversion (or dissipation) of magnetic into kinetic energy. It has been argued that magnetic reconnection could be an important mechanism to allow such conversion \citep[e.g.,][and references there in]{giannios_10,dalpino_kowal_15,dalpino_etal_2018,werner_etal_2018,giannios_uzdensky2019}. 
Magnetic reconnection is, in fact, now regarded as a fundamental plasma process in a variety of flaring phenomena in the Universe. Not only in the solar system in association to solar flares and the Earth magnetotail storms, where reconnection has been directly observed, but also beyond that.  Reconnection is now invoked  to explain energy dissipation and specially high energy non-thermal variable emission in systems like pulsar wind nebulae \citep[e.g.,][]{lyubarsky_etal_2001,clausen-brown_2012,cerutti_etal_2014}, jets and accretion disks in microquasars and AGNs\citep{dalpino_lazarian_2005,giannios_etal_2009,dalpino_etal_2010,nalewajko_etal_2011,
giannios_10,giannios_2011,mckinney_2012,kadowaki_etal_15,singh_etal_15,khiali_etal_15,
singh_etal_16,sironi_etal_2015,dalpino_etal_2018,kadowaki_etal_18,kadowaki_etal_2019,
christie_etal_19,fowler_etal_2019,nishikawa_etal_2020}, and gamma-ray bursts (GRBs) \citep[e.g.,][]{drenkhahn_2002,giannios_spruit_2007, zhang_yan_11, mckinney_2012}.

Magnetic reconnection occurs when two magnetic fluxes of opposite polarity encounter, then partially break 
and rearrange their configuration 
at a velocity $V_{rec}$, which is a substantial 
fraction of the local Alfv\'{e}n speed if reconnection is fast \citep[e.g.][]{priest_etal_03,zweibel_yamada_2009,lazarian20}.
Different processes such as plasma instabilities, anomalous resistivity, and turbulence, may trigger fast reconnection. The latter process, in particular, is very efficient and probably the main driving mechanism of fast reconnection in collisional astrophysical flows. 
Since turbulence causes 
an efficient
field-fluid slippage and stochasticity of the magnetic field lines, bringing initially distant lines into close separations through Richardson diffusion \citep[][]{eyink_etal_2013, jafari_etal_2020}, this leads to many patches reconnecting simultaneously, making the reconnection rate very fast \citep[and actually independent of the intrinsic plasma microscopic magnetic resistivity; see, e.g., ][]{lazarian_vishiniac_99,eyink_aluie_2006, kowal_etal_09,eyink_etal_2011, eyink2015, takamoto_etal_15}.
The  rearrangement of the magnetic field configuration may convert  magnetic into thermal and kinetic energies \citep[e.g.][]{yamada_etal2016}, and allows for efficient particle acceleration in a Fermi-like stochastic process \citep{dalpino_lazarian_2005} \citep[see also  reviews on this process, e.g. in][]{dalpino_kowal_15,matthews_etal_2020}. Particle acceleration in reconnection regions has been successfully tested in several numerical works employing multi-dimensional (2D and 3D)  magnetohydrodynamic (MHD) simulations with test particles \citep[e.g., ][]{kowal_etal_2011,kowal_etal_2012,delvalle_etal_16,beresnyak_etal_2016, dalpino_etal_2018,dalpino_etal_2019}, and particle-in-cell (PIC) simulations \citep[e.g.][]{drake_etal_2006,zenitani_H_2007,zenitani_H_2008,lyubarsky_etal_2008,drake_etal_2010,clausen-brown_2012,cerutti_etal_2012,sironi_spitkovsky_2014,li_etal_2015,lyutikov_etal_2017,guo_etal_2015,guo_etal_2016,werner_etal_2018,werner_etal_2019}.

As remarked, this mechanism can operate in magnetized astrophysical flows in general, and particularly in relativistic jets, in the regions near the jet base where they are possibly magnetically dominated \citep[see, e.g.,][]{zamaninasab_etal_14}. To explore reconnection in these objects is the main focus of this work. 
Magnetic reconnection seems to be specially helpful to solve  puzzles related to the very high energy (VHE) emission of blazar jets. Blazars are AGNs with highly beamed relativistic jets that point closely to the line of sight \citep{Blandford_Rees_1978,urry_padovani_1995}. They are the most common extragalactic sources of $\gamma$-rays, both in GeV \citep[e.g.,][]{acero_etal_2015} and TeV  bands \citep[e.g.,][]{Rieger_etal_2013,tavecchio_2020}. Their powerful non-thermal emission, spanning the entire electromagnetic spectrum has power-law distribution function with variability time-scales ranging from days \citep[as in Mrk 421,][]{kushwaha_etal_17}, to minutes \citep[as, e.g., in PKS 2155-304, Mrk 501, 3C 279, and 3C 54.3; see][]{aharonian_etal_07,albert_etal_07,ackermann_etal_2016,britto_elal_2016}. This multi-wavelength emission is usually attributed to relativistic particles accelerated stochastically in recollimation shocks along the jet and in their head \citep[e.g.,][]{mizuno_etal_15}. However, these shocks may be not powerful enough to probe emission in the magnetically dominated regions of these jets \citep[e.g.][]{sironi_elal_2013,dalpino_kowal_15, bell_etal_2018}. Striking examples are some of the blazars above with very short duration TeV flares (of a few minutes only). These imply explosive and extremely compact acceleration/emission regions ($< R_S/c$) with Lorentz factors much larger than the typical jet bulk values (which are $\Gamma \simeq 5 – 10$) in order to avoid electron-positron pair creation and thus, entire gamma-ray absorption within the source \citep[e.g.,][]{begelman_etal_08}. A strong candidate mechanism (and possibly the only one) that seems to be able to circumvent this problem and explain this high variability and compactness of the TeV emission is fast magnetic reconnection involving misaligned current sheets inside the jet \citep[see, e.g.][]{giannios_etal_2009,giannios_2013,kushwaha_etal_17}. A similar process has been also proposed for gamma-ray bursts \citep[GRBs; e.g.,][]{giannios_2008,zhang_yan_11}.

Instabilities occurring in the jet can drive turbulence and thus, 
as described above, to substantial randomness and diffusion of the  field lines,
leading to the production of current sheets where fast turbulent reconnection is triggered \citep[e.g.,][]{spruit_etal_2001,giannios_spruit_2006,singh_etal_16,Tchekhovskoy_etal_2016,barniol_eltal_2017,gill_etal_2018,dalpino_etal_2018,nishikawa_etal_2020}.
Jets with helical magnetic field structure, in particular, are susceptible to current-driven kink (CDK) instability \citep{begelman_1998,giannios_spruit_2006,mizuno_etal_09,mizuno_etal_11,mizuno_etal_12,mizuno_etal_14,alves_etal_2018,das_begelman2019}.  
MHD simulations of the launching and propagation of relativistic jets indicate that this instability can be triggered where the jet recollimates \citep{bromberg_etal_2016, nishikawa_etal_2020}. 
Several concomitant works \citep[see][]{porth_etal_2015,singh_etal_16,Tchekhovskoy_etal_2016, striani_etal_2016}  have also revealed that this instability can operate in the jet spine over limited regions without disrupting the entire jet structure, which is compatible with the observations that these jets can propagate over very large distances and remain stable. Besides, these works have also confirmed that CDK instability converts substantial magnetic into kinetic energy and also drives magnetic reconnection. In particular, the three-dimensional special relativistic magnetohydrodynamic (3D SRMHD) simulations of rotating, Poynting flux dominated jets with helical fields of \citet*{ singh_etal_16} have identified CDK instability induced turbulence and the formation of current-sheets with fast reconnection rates $\sim  0.05V_A$. 

Despite the extensive study of CDK, none of these previous works have performed  a detailed and systematic identification of the location of all the fast reconnecting current-sheets driven by the CDK instability over the entire jet domain, nor a precise determination of the reconnection rates, or the magnetic power released in these sites. This determination is very important particularly because these are the potential sites for particle acceleration and non-thermal emission, as stressed above. In this work, we expand upon the work of \citet*{singh_etal_16}, and perfom a systematic analysis of  reconnection sites inside the simulated jet, in a similar way to that employed before in MHD simulations of turbulent environments \citep{zhdankin_etal_13}, in shearing-box accretion flows \citep*{kadowaki_etal_18}, or turbulent current-sheets with a slab geometry \citep{kowal_etal_09,kowal_etal_2020}. To this aim, we employ here a modified version of the magnetic reconnection search-algorithm developed in \citet{kadowaki_etal_18} to incorporate relativistic effects. We apply this search-algorithm to several snapshots of the simulated jet which allow us to obtain robust averages of the reconnection rates and the magnetic power of multiple reconnection events, as well as the time variability pattern of them. As an example to show the robustness of the method, we scale our results to the blazar jet Mrk 421 \citep{kushwaha_etal_17} and find that the magnetic reconnection power is compatible with the observed gamma-ray power, and the time variability driven by reconnection is consistent with the observed flaring in $\gamma-$rays. We should remark that, in order to obtain a full understanding of how this magnetic energy released by the CDK instability is channeled into energetic nonthermal particles, in a companion paper we have shown results of particle acceleration, injecting thousands of test particles in this 3D SRMHD jet simulation \citep[][hearafter \citetalias{medina_torrejon_etal2020}]{medina_torrejon_etal2020}.
Preliminary results of this study have been presented in \citet{dalpino_etal_2018}, \citet{kadowaki_etal_2019}, and \citet{dalpino_etal_2019}.

The paper is organized as follows, in Section \ref{sec:num_method}, we describe the numerical methodology. In Section \ref{sec:rec} we present our results. In Section \ref{sec:application}, we show the applications of our results to Mrk 421 source. Finally in Section \ref{sec:conclusions}, we draw  our conclusions and discuss our findings.

\section{Numerical Method}
\label{sec:num_method}

The numerical solution of a Poynting-flux dominated jet under the action of the current-driven kink instability has been obtained from the ideal special relativistic magnetohydrodynamic (SRMHD) equations, that describe the macroscopic behavior of a relativistic magnetized fluid. To this aim, we have used the \texttt{RAISHIN} code based on a 3+1 formalism of the general relativistic (GR) conservation laws of continuity, energy-momentum, and Maxwell's equations in a curved spacetime \citep[see][]{mizuno_etal_06}. In the present work, we have adopted an ideal gas equation of state with $p= (\Gamma - 1) \rho e$, where $p$ is the thermal pressure, $\rho$ the rest-mass density, $\rho e$ the specific internal energy density, and $\Gamma = 5/3$ the adiabatic index. We have also adopted the flat Minkowski spacetime as an appropriate metric to perform the simulations in the special relativistic regime. 

We have employed a Harten-Lax-van Leer-Einfeldt (HLLE) approximate Riemann solver to compute the intercell fluxes of the computational grid \citep[see][]{einfeldt_88}, a flux-interpolated constrained transport method to maintain a divergence-free condition for the magnetic field \citep[flux-CT,][]{toth_00}, and a third-order Runge-Kutta scheme to advance the equations in time. Furthermore, we have used a second-order MC slope-limiter for the reconstruction step, and since the \texttt{RAISHIN} code uses a conservative scheme, an inversion procedure based on the method of \citet{mignone_mcKinney_07} have been used to transform the conserved variables into the primitive ones \citep[for more details, see also][]{mizuno_etal_06,mizuno_etal_11}. Finally, the equations have been solved in dimensionless code units (c.u.) without a factor of $4\pi$ and considering the light speed as the velocity unit (i.e., $c=1$)\footnote{We will hereafter omit the notation ``c.u.'' for simplicity, except when the scale-free (code) units are converted into physical units (see more details in Section \ref{sec:phunit_lightcurve}).}.

\subsection{Initial Conditions}

Following the work of \citet{mizuno_etal_12} \citepalias[see also][]{medina_torrejon_etal2020}, we have used as initial condition a force-free helical magnetic field profile that decreases as a function of radius and with constant pitch ($\mathcal{P}=RB_z B_\phi$). In a cylindrical coordinate system ($R$, $\phi$, $z$),  these equations are given by:

\begin{equation}
B_z=\frac{B_0}{1+(R/R_0)^2} ~\textrm{and}
\end{equation}
\begin{equation}
B_{\phi}=-\frac{B_0(R/R_0)[1+(\Omega R_0)^2]^{1/2}}{1+(R/R_0)^2} ~,     
\end{equation}
where $B_0=0.7$ is the magnetic field amplitude, and $R_0=0.25$ the jet core radius (all in code units). The angular velocity of the jet is given by:
\begin{equation}
\Omega=\left\{\begin{matrix}
\Omega_0 & \mathrm{if}\,\,\, R\leq R_0 \\ 
\Omega_0(R_0/R) & \mathrm{if}\,\, R>R_0
\end{matrix}\right. ~,
\end{equation}
where $\Omega_0=2.0$ is the angular velocity amplitude.

We have also used a decreasing rest-mass density and pressure profiles with radius given by:
\begin{equation}
\rho=\rho_1 \sqrt{B^2\over{B^2_0}} ~\textrm{and}    
\end{equation}
\begin{equation}
p=\left\{\begin{matrix}
p_0 & \mathrm{if}\,\, R\leq R_p \\ 
p_0(R_p/R) & \mathrm{if}\,\, R>R_p
\end{matrix}\right. ~,
\end{equation}
where $\rho_1=0.8$ is the density at the jet spine ($R=0$), and $p=0.02$ is a constant pressure inside a characteristic radius $R_p=0.5$, which is used to keep a non-dominant thermal pressure force in the initial force-free configuration.

The initial conditions for the vertical and azimuthal components of the velocity field have been defined from the drift velocity $\boldsymbol{v} = c\boldsymbol{E} \times \boldsymbol{B}/B^2$ and are given by:
\begin{equation}
v_z=-\frac{B_\phi B_z}{B^2}\Omega R ~\textrm{and}
\label{eq:vz}
\end{equation}

\begin{equation}
v_\phi=\left(1-\frac{B_\phi^2}{B^2}\right )\Omega R ~.
\label{eq:vphi}
\end{equation}
Finally, an unstable perturbed profile has been used in the radial velocity component
to trigger the CDK instability \citep[see, e.g.,][]{mizuno_etal_11, mizuno_etal_12}, so that
\begin{equation}
\label{eq:vr}
v_R=\frac{\delta v}{N}\exp{\left ( -\frac{R}{R_0} \right )}\sum_{n=1}^{N}\cos(m\phi)\sin\left ( \frac{\pi n z}{L_z} \right ) ,
\end{equation}
where $\delta v = 0.01$ is the perturbation amplitude, $N=8$ is the total number of
wavelengths used to excite the kink mode ($m=1$) in the system, and $L_z=6$ is the size of the computational domain in the $z$ direction.

\subsection{Computational Domain and Boundary Conditions}

We have used a Cartesian grid ($x$,$y$,$z$) considering a computational domain of size $6 \times 6 \times 6$ (code units). This coordinate system is particularly appropriate to the study and identification of the magnetic reconnection sites employing the algorithm developed in \citet{kadowaki_etal_18} (see more details in Section \ref{sec:algorithm}). We have also applied standard outflow boundary conditions in the $x$ and $y$ directions (i.e., zero gradients for all variables), which  are different from the fixed boundaries used in \citet{mizuno_etal_12}. Even though the latter case maintains the stability of the jet rotation at the edges of the computational box, outflow conditions allow all the variables to evolve freely at the transverse boundaries avoiding artificial numerical effects. Finally, as in \citet{mizuno_etal_09,mizuno_etal_11,mizuno_etal_12},  periodic boundaries have been applied in the $z$-direction to maintain the CDK instability growing until the saturation and disruption inside of the domain (see Section \ref{sec:rec}).

\subsection{Magnetic Reconnection Search Algorithm and Reference Frames}
\label{sec:algorithm}

\begin{figure*}
  \centering
  \includegraphics[scale=0.5]{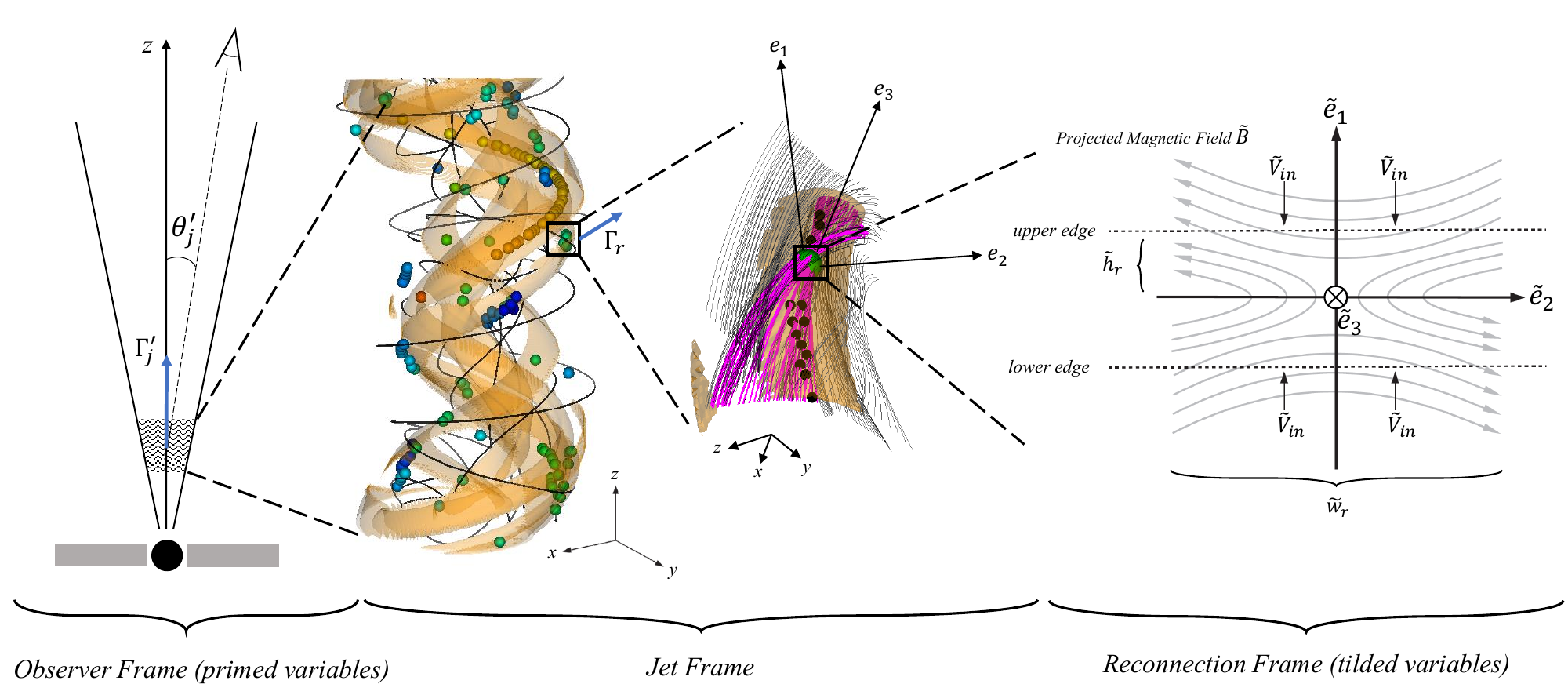}
  \caption{The schematic diagram shows different structures (and frames) analyzed in the present work. The first sketch (from the left to the right) shows the relativistic jet at the observer reference frame (primed variables). The hatched region corresponds to the simulation domain moving with an angle $\theta_j^{\prime}$ to the line of sight and with a bulk Lorentz factor $\Gamma_j^{\prime}$. The second and third figures show the evolved structures of the simulation at the jet reference frame. The colored circles correspond to reconnection sites and the streamlines correspond to the magnetic field. Finally, the last sketch \citep[adapted from][]{kadowaki_etal_18} depicts the details of a reconnection region in a coordinate system at the reconnection reference frame (tilded variables).}
  \label{fig:scheme_jet}
\end{figure*}

In order to search magnetic reconnection events in our Poynting-flux dominated jet simulations in the relativistic regime, we have used a modified version of the algorithm developed by \citet{kadowaki_etal_18} \citep[see also,][]{zhdankin_etal_13}. As in the previous work, we select a sample of cells with a current density value ($\boldsymbol{J}=\boldsymbol{\nabla}\times \boldsymbol{B}$) five times higher than the average one taken in the whole system ($|\boldsymbol{J}_{max}|> \epsilon \langle |\boldsymbol{J}| \rangle$, for $\epsilon=5$), and choose those cells where $|\boldsymbol{J}_{max}|$ is a local maximum within a subarray data cube of size $3 \times 3 \times 3$ cells. Then, we  evaluate the eigenvalues and eigenvectors of the current density Hessian matrix around each local maximum cell to obtain a new local coordinate system ($e_1$,$e_2$,$e_3$; see Figure \ref{fig:scheme_jet}) centered in the magnetic reconnection site since it may not be necessarily aligned with one of the axes of the Cartesian coordinate system \citep[for more details; see][]{kadowaki_etal_18,zhdankin_etal_13}. The edges of the reconnection (or diffusion) regions are defined when the current density decays to half of the maximum value along the $e_1$, $e_2$ and $e_3$ directions ($|\boldsymbol{J}_{edge}|=0.5|\boldsymbol{J}_{max}|$).

These latter steps are applied in the coordinate frame (hereafter jet frame), where the relativistic jet simulations are performed (see the previous sections). However, since each local maximum cell can move relative to the jet frame with relativistic velocities, we have introduced the reconnection comoving frame (tilded variables) to evaluate quantitatively the reconnection events.  We then transformed all the variables inside each subarray data cube from the jet  to the reconnection frame, via generalized Lorentz transformations. In order to remove false-positive events, we also introduced additional criteria for the magnetic and velocity fields (at the reconnection frame), where we selected only those regions with opposite magnetic field and inflow velocity components at the upper and lower edges of the reconnection sites. Furthermore, all the identified regions in the final sample have at least one outflow velocity component at the left or right edges (see more details in Section \ref{sec:edges}).

With this algorithm, we can evaluate several features of the diffusion regions, such as the reconnection structures, the reconnected components of the magnetic field, and the magnetic reconnection rate, showing the advantages of this analysis. In  Section \ref{sec:rec}, we will focus on the study of these features.

Finally, besides the jet and reconnection reference frames mentioned above, we will also have to  deal with the observer's frame (primed variables). We assume that the entire computational domain is moving relative to this frame with a bulk Lorentz factor $\Gamma_j^{\prime}$, and an angle $\theta_j^{\prime}$ to the line of sight (see Figure \ref{fig:scheme_jet} for a schematic view of the scenario studied in the present work). This is reasonable, since the initial vertical and azimuthal velocity components (equations \ref{eq:vz} and \ref{eq:vphi}, for $\Omega_0=2.0$) are mildly relativistic and defined by the drift velocity \citep[similar assumption was used by][]{mizuno_etal_12}. In this work, the observer's frame will be used to build a synthetic light curve based on the observations of the Blazar Mrk 421, considering  reconnection events as the primary source for dissipation and variable emission at high energies (see more details in Section \ref{sec:application}).

\subsection{Simulation and Algorithm Parameters}

As mentioned previously, the simulations were performed using a computational domain of size $6 \times 6 \times 6$ (c.u.), so that a small portion of the magnetized relativistic jet can be reproduced. Besides, we also considered three different resolutions ($120^3$, $240^3$, and $480^3$ cells) for convergence test purposes. For the reconnection search algorithm parameters, we identify cells with $|\boldsymbol{J}_{max}|> \epsilon \langle |\boldsymbol{J}| \rangle$ (for $\epsilon=5$), and consider a subarray data cube of size $3\times3\times3$ cells to select only the local maximum current densities. Finally, the diffusion region's edges were defined by a current density criterion $|\boldsymbol{J}_{edge}|=0.5|\boldsymbol{J}_{max}|$ for our reference model (m240ep0.5). However, in addition to this, we have also considered extended diffusion regions with $0.1|\boldsymbol{J}_{max}|$ and $0.05|\boldsymbol{J}_{max}|$. The parameters of the simulations are shown in Table \ref{tab:parameters}, and each model name is composed of the resolution (m$120$, m$240$, and m$480$) plus the edge position values (ep$0.5$, ep$0.1$, and ep$0.05$).

\input{table00.tex}

\section{Numerical Results}
\label{sec:rec}

\citet{singh_etal_16} have performed 3D SRMHD simulations of Poynting-flux dominated relativistic jets with an initial helical magnetic field structure suitable for jets near the launching region \citep[see also][]{mizuno_etal_09,mizuno_etal_11,mizuno_etal_12,mizuno_etal_14}. Considering models with moderate magnetization ($\sigma \sim 0.1-1$), and different density profiles of the jet and the environment, the authors induced small precession perturbations to trigger the CDK instability that in turn excites turbulent fast magnetic reconnection, as stressed in section \ref{sec:intro}. Investigating regions of maximum current density in the jet domain, they  derived a reconnection rate of $\sim 0.05 V_A$ \citep[see also][]{takamoto_etal_15}. Extending  the work of \citet{singh_etal_16}, we employ here the method developed by \citet{kadowaki_etal_18}, described in Section \ref{sec:algorithm}, to search and quantify every reconnection events in these relativistic jets. 

\begin{figure*}
  \begin{center}
    \includegraphics[scale=0.51]{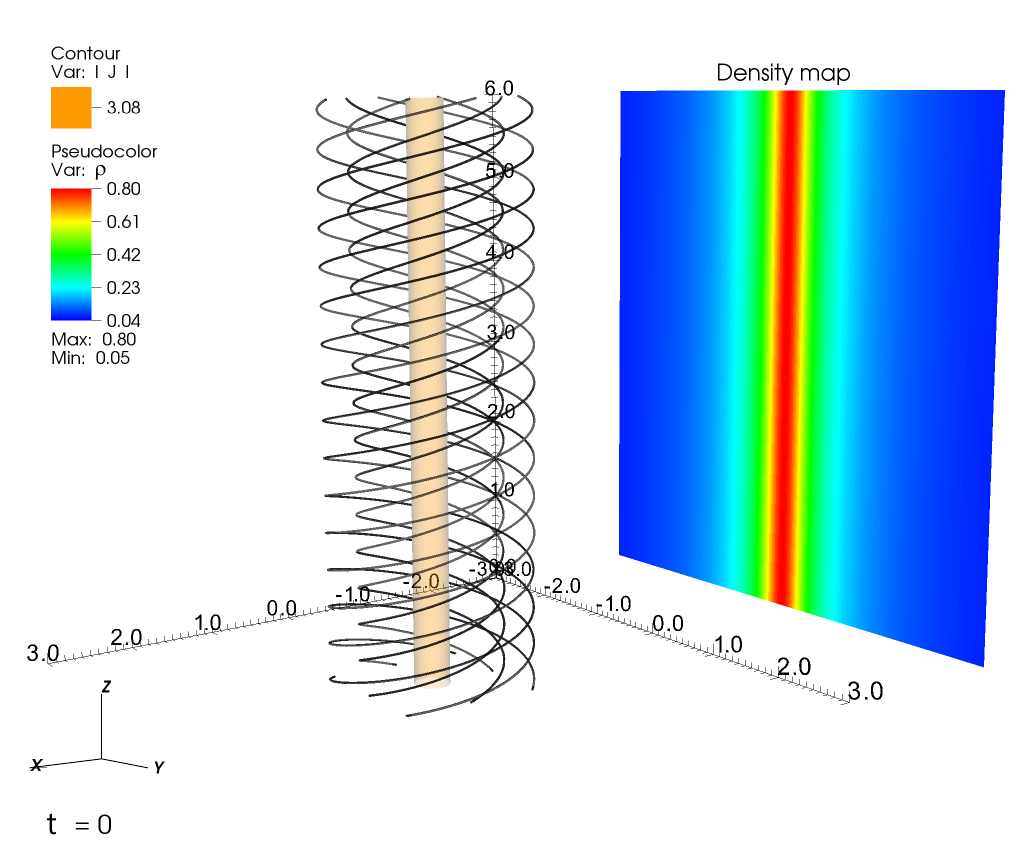}
	\includegraphics[scale=0.51]{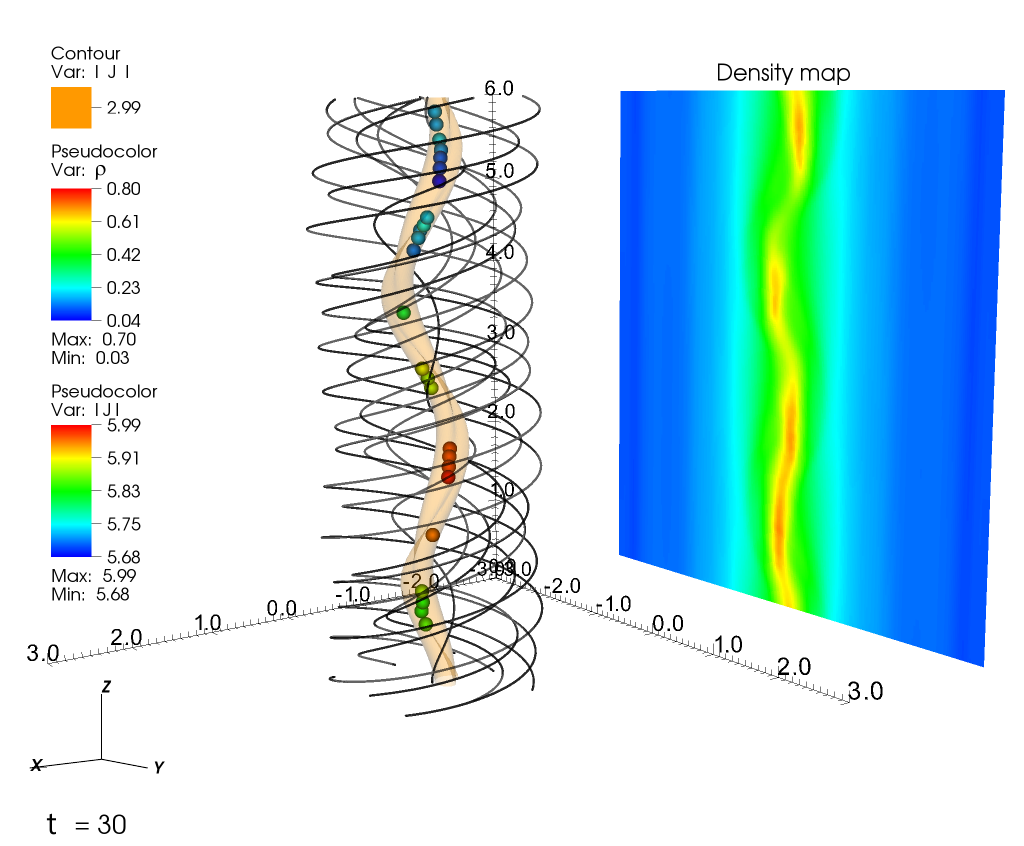}
	\includegraphics[scale=0.51]{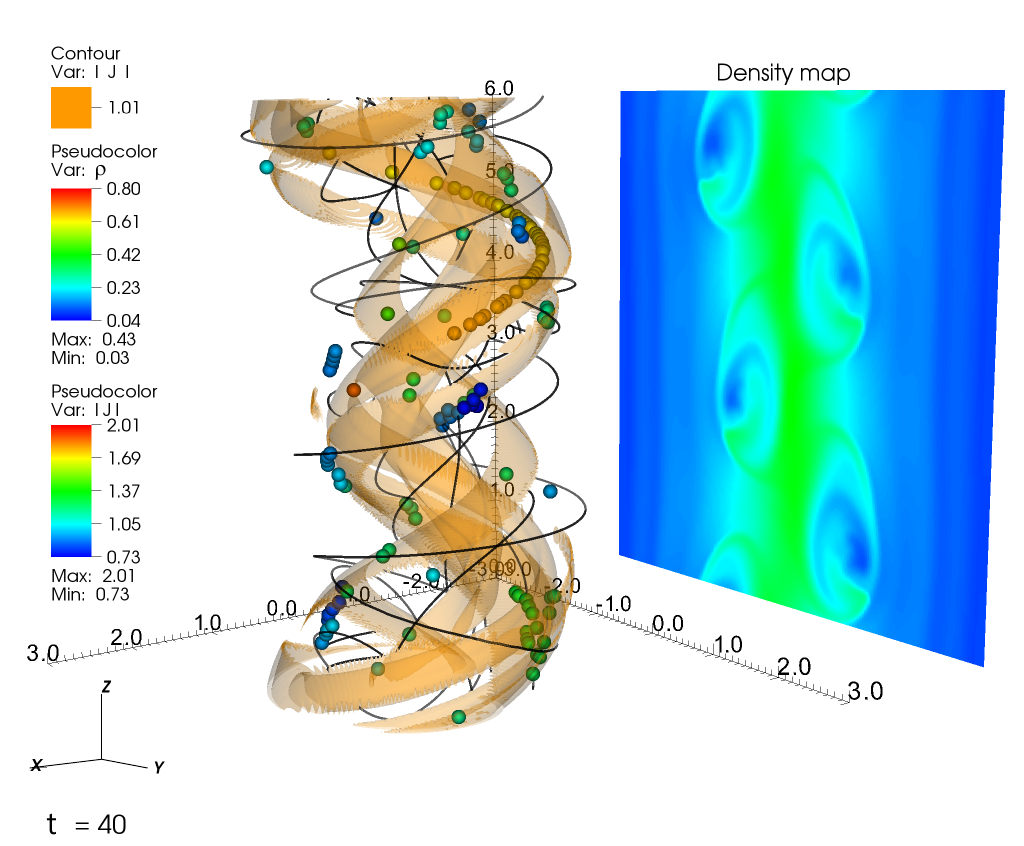}
	\includegraphics[scale=0.51]{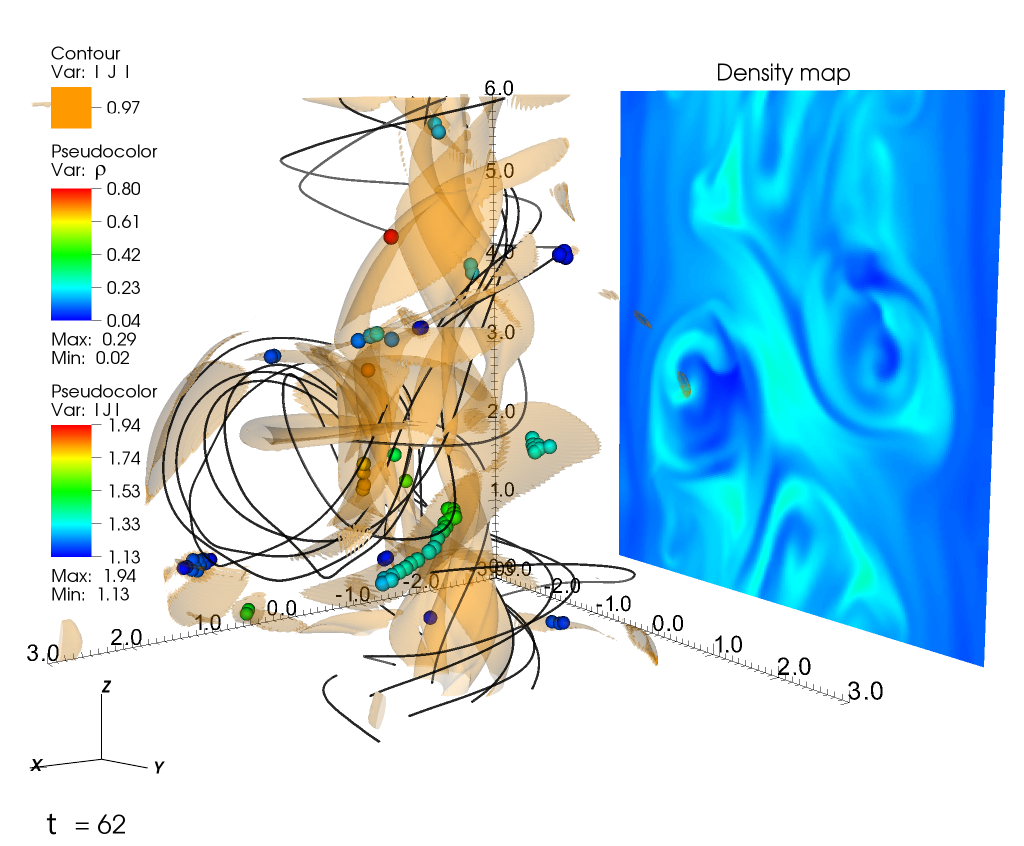}
	\caption{Time evolution of the jet at $t=0$, $30$, $40$ and $62$ c.u. (from top left to bottom right). The diagrams show isosurfaces of half of the maximum current density intensity $|\boldsymbol{J}|$ (orange color), the magnetic field topology (black lines), the density profile at the middle of the box ($y-z$ plane at $x=0$), and the magnetic reconnection sites identified by the search algorithm in the jet frame (colored circles along the jet correspond to different current density magnitudes in code units).}
	\label{fig:jetevol}
  \end{center}
\end{figure*}

We start our study considering the reference model m240ep0.5 (see Table \ref{tab:parameters}). The diagrams of Figure \ref{fig:jetevol} show the current density isosurfaces of half of the maximum $|\boldsymbol{J}|$ (orange color), the magnetic field topology (black lines), the density profile at the middle of the box ($y-z$ plane at $x=0$), and the magnetic reconnection sites (or diffusion regions) identified by the algorithm (represented by colored circles along the distorted jet spine by the CDK instability).  The different 
colors of the circles correspond to different values of the current density magnitude at $t=0$, $30$, $40$ and $62$ (all in code units) in the jet frame. 

We note that there are some differences between the simulations presented here and those by \citet{singh_etal_16}, including the reduced size of the box and the use of periodic boundaries in the z-direction, rather than outflow boundaries, as in  \citet{mizuno_etal_12}. This implies that the CDK instability is continuously driven in the system, rather than propagating downstream and eventually leaving the system as seen in \citet{singh_etal_16}. Therefore, as the instability grows and magnetic energy is transformed into kinetic energy, the amplitude of the helical distortions of the jet spine increases up to a complete disruption of the jet, as we see at $t=60$ in Figure \ref{fig:jetevol}. In \citetalias{medina_torrejon_etal2020} (see their Figure 2), we analysed in detail the evolution of the CDK instability and find that its non-linear growth starts around $t \sim 30$ and saturates with the formation of a plateau in the kinetic energy density around $t\sim 40$. The instability drives turbulence which develops completely after this time inducing magnetic reconnection and the formation of several current sheets. Figure \ref{fig:jetevol} shows that even during the growth of the CDK instability, before saturation, at $t=30$, there are already sites of reconnection \citepalias[see also][]{medina_torrejon_etal2020}. The analysis performed in the next sections will show that only part of them are real locci of fast reconnection.
Furthermore, in the same snapshot $t=30$, despite the high value of $|\boldsymbol{J}|$ for each site, the associated current density isosurfaces show thick structures implying that the magnetic field lines are not as much accumulated. This reflects in a low average reconnection rate as we will see in section \ref{sec:vrec}. In $t=40$ and beyond, on the other hand, there are several reconnection sites following the growing wiggling amplitude all along the jet spine, and the associated current density isosurfaces become thinner, characterizing fast reconnection current sheets (section \ref{sec:vrec}). After $t=62$, the helical spine structure is completely disrupted, and the maximum value of the mass density reduces to $\sim 36\%$ of its initial value.

In all diagrams of Figure \ref{fig:jetevol}, several reconnection sites close to each other were identified by the algorithm, due to our choice of the size of the cubic subarray of data (of $3 \times 3 \times 3$ cells; see Section \ref{sec:algorithm}). With this choice, it has been possible to obtain not only the primary site (with the maximum value of $|\boldsymbol{J}|$), but also the secondary ones along the current sheet produced by the encounter of magnetic fields of opposite polarities. Due to this approach, we can see in the diagrams, as time evolves, the fragmentation of the large-scale current isosurfaces (and the diffusion regions) that can be associated with turbulence and fast magnetic reconnection events (after $t=40$) as we will see in the subsequent sections. Similar behavior was obtained in resistive MHD simulations of braided coronal loops conducted by \citet{pontin_etal_11}, where the magnetic reconnection leads to a cascade of multiple small-scale events with turbulent-like behaviour \citep[see also][]{kowal_etal_19}, which is in agreement with the predictions of the turbulent magnetic reconnection mechanism \citep[][]{lazarian_vishiniac_99, eyink_etal_2011,santos-lima_etal2020}.

\subsection{Magnetic Reconnection Structures}
\label{sec:mag_structures}

\begin{figure*}
  \begin{center}
    \includegraphics[scale=0.582]{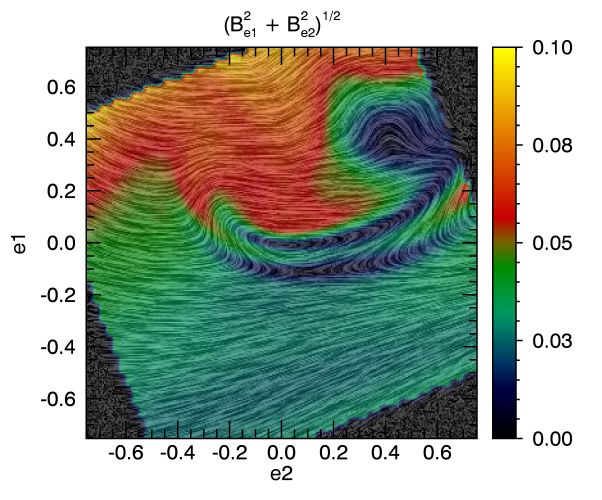}
	\includegraphics[scale=0.582]{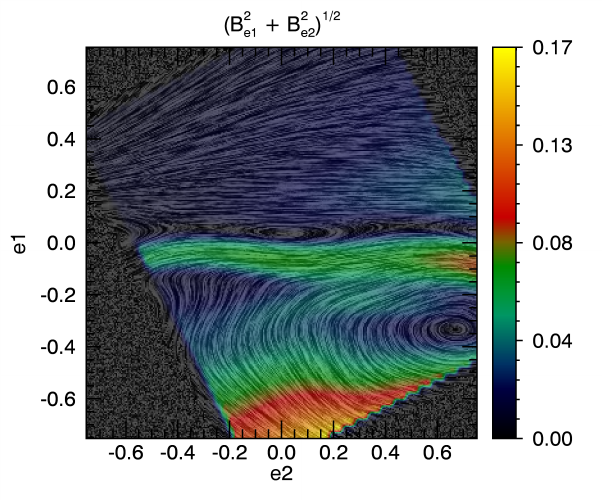}
	\includegraphics[scale=0.582]{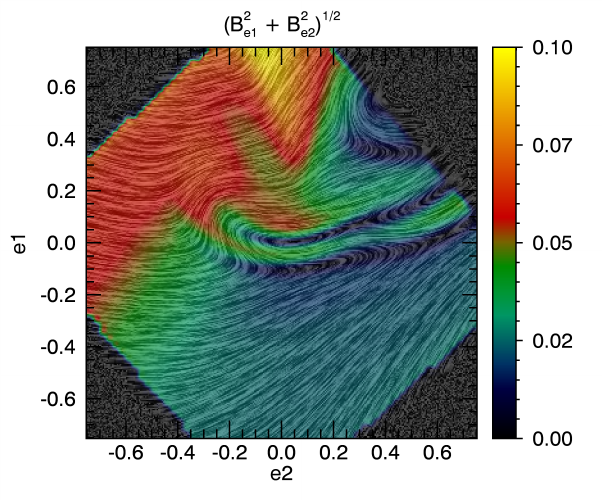}
	\\
    \includegraphics[scale=0.582]{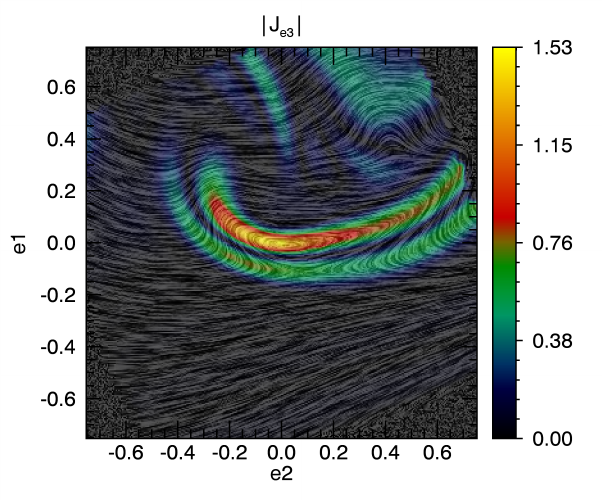}
    \includegraphics[scale=0.582]{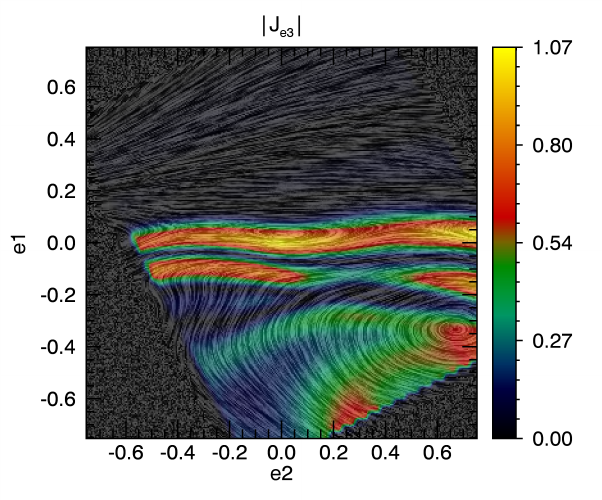}
    \includegraphics[scale=0.582]{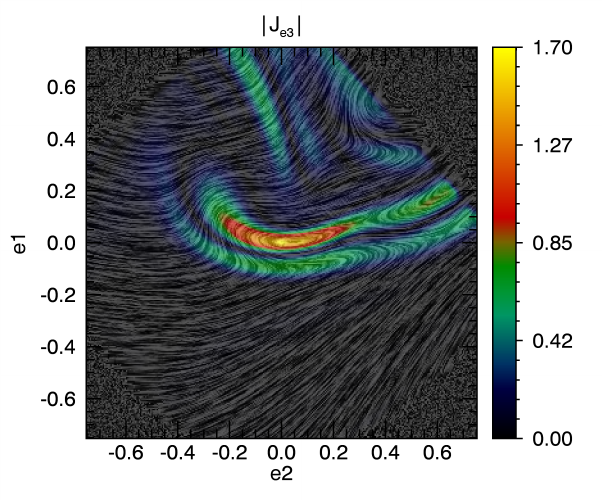}
	\caption{Diagrams showing in detail three magnetic reconnection sites in the jet frame, selected from snapshot $t=50$. The diagrams were produced by a line integral convolution (LIC) method combined with the $2D$ projection of the magnetic field (top diagrams) and current density (bottom diagrams) magnitudes.}
	\label{fig:sample_lic}
  \end{center}
\end{figure*}

We have analyzed the magnetic reconnection structures for a sample of sites identified by the search algorithm in order to verify how reconnection events can behave along the relativistic jet. Figure \ref{fig:sample_lic} shows three candidate sites in the jet frame, selected at $t=50$. The $2D$ maps have been produced using the line integral convolution (LIC) method \citep{cabral_leedom_93} to represent, at the same diagram, the magnetic streamlines, and the magnetic field (top diagrams) and current density (bottom diagrams) magnitudes. The diagrams were obtained by a cubic interpolation of the data (for visualization purposes) and correspond to an arbitrary slice in the $e_{1}-e_{2}$ plane (the local coordinate system at the jet frame, see Section \ref{sec:algorithm}). The diagrams show clearly the anti-correlation between the magnetic field and the current density, as we should expect for reconnection events. Besides, the bottom diagrams show the high concentration of the magnetic field around the reconnection sheets. The first column from left shows an elongated ``bow-shaped'' magnetic island with an X-point-like structure right below, whereas the middle diagrams show a complex topology with at least three magnetic islands. The third column shows similar behavior with two islands (already reconnected) separated by an X-point-like structure and accumulated magnetic field lines (in  reconnection) below them. 
This study is useful to follow the time evolution of these structures since each example may represent reconnection events at different stages. Moreover, such analysis allows us to identify false-positives events in our sample (i.e., regions identified  by the search algorithm that are actually not  associated with reconnection events; see section \ref{sec:algorithm}). However, for  a complete analysis, it would be necessary to check about $200$ sites per snapshot separately (in a total of $660$ snapshots for model m240ep0.5), but that is out of the scope of the present study. For future works, machine learning techniques can be applied to recognize and select automatically such events from these diagrams.

\begin{figure}
  \begin{center}
    \includegraphics[scale=0.89]{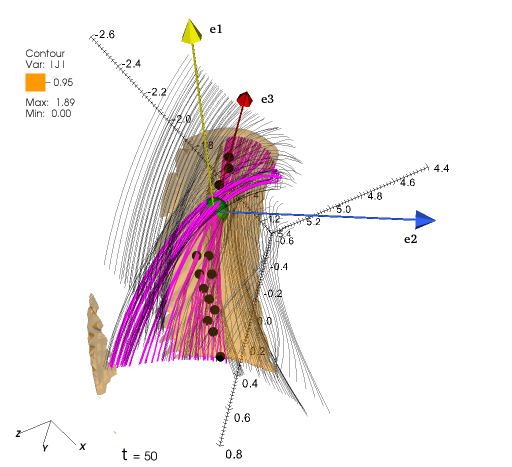}\\
	\includegraphics[scale=0.69]{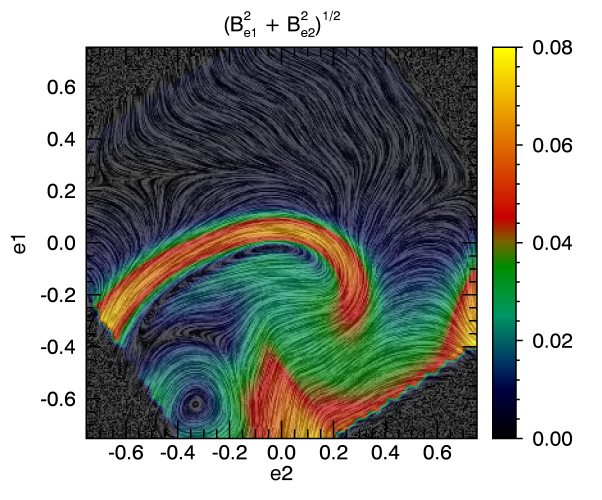}\\
    \includegraphics[scale=0.69]{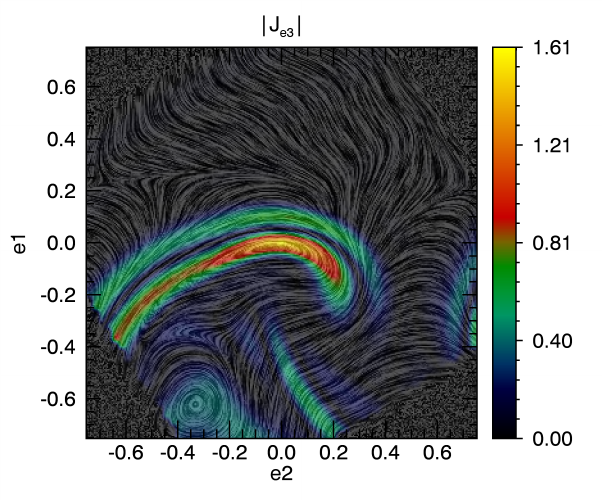}
	\caption{The top diagram shows a zoom-in plot of the magnetic field streamlines (black and magenta lines) around a sample of identified reconnection sites (green and black circles), at $t=50$, in the jet frame. The colored arrows represent the axes of the local coordinate system, and the orange isosurface corresponds to the associated current sheet with half of the maximum value of $|\boldsymbol{J}|$ at the primary reconnection site (green circle). The middle and bottom diagrams show the corresponding $2D$ LIC maps (as in Figure \ref{fig:sample_lic}).}
	\label{fig:3drec}
  \end{center}
\end{figure}

Despite the information provided by the LIC diagrams (Figure \ref{fig:sample_lic}), the $2D$ projection of the magnetic field components is not enough to reveal the complex $3D$ topology of a reconnection event. Besides, $3D$ reconnection does not always imply that the three components of the magnetic field are annihilating at a time as we see several cases in our work, where only one of the components of the magnetic field is annihilated \citep[see also, e.g.,][and references therein]{priest_etal_03, parnell_etal_10, yamada_etal_10, pontin_etal_11}. The top diagram of Figure \ref{fig:3drec} shows a zoom-in plot of the magnetic field streamlines (black and magenta lines) around a sample of identified reconnection sites (green and black circles) in the jet frame at $t=50$. The colored arrows represent the axes $e_{1}$, $e_{2}$, and $e_{3}$ (yellow, blue, and red) of the local coordinate system, the orange isosurface corresponds to the associated current sheet with half of the maximum value of $|\boldsymbol{J}|$ at the primary reconnection site (green circle), and the black circles correspond to secondary reconnection events. The black lines represent the asymptotic (non-reconnected) magnetic field far away from the diffusion region whereas the magenta lines represent the twisted and braided magnetic field lines that produce a thin and strong current density isosurface (as in the diagrams of Figure \ref{fig:jetevol}, after $t=40$) with the maximum value at the green circle's position. As we expected, the $e_{1}$-axis is perpendicular to the current sheet, whereas the $e_{2}$-$e_{3}$ plane is aligned with it. The $e_{3}$-axis matches the direction of the local magnetic guide field (and the current sheet), proving the efficiency of the algorithm in separating the dominant non-reconnected magnetic component from those in reconnection. 

Finally, the $2D$ LIC maps at the middle and bottom of Figure \ref{fig:3drec} show the magnetic and current density magnitudes (as in Figure \ref{fig:sample_lic}) around the primary reconnection site (green circle in the top diagram). Both maps show a magnetic island topology as a result of the projection of the reconnected components in the $e_{1}-e_{2}$ plane. The local magnetic guide field is still present, but hidden by the projection. Therefore, despite the useful information obtained from these $2D$ maps, the overall $3D$ scenario and the real nature of the magnetic islands, which are $2D$ projections of reconnected flux tubes
\citep{kowal_etal_2011,kowal_etal_2012}, should be analyzed carefully. This discussion is important since the formation of current sheets along the jet due to reconnection events is expected \citep[see, e.g.,][]{giannios_etal_09,christie_etal_19}, but the real magnetic topology can be far more complex than island-like structures. 

\subsection{Profiles at the Edges of the Diffusion Regions}
\label{sec:edges}

In the previous section, we have shown the capability of the search algorithm to recognize magnetic reconnection sites and described qualitatively the $2D$ and $3D$ topologies of such events along the relativistic jet. In this section and the next ones, we will present a quantitative analysis of the diffusion region in the reconnection reference frame (tilded quantities), as described in Section \ref{sec:algorithm}.

\begin{figure*}
  \includegraphics[scale=0.4]{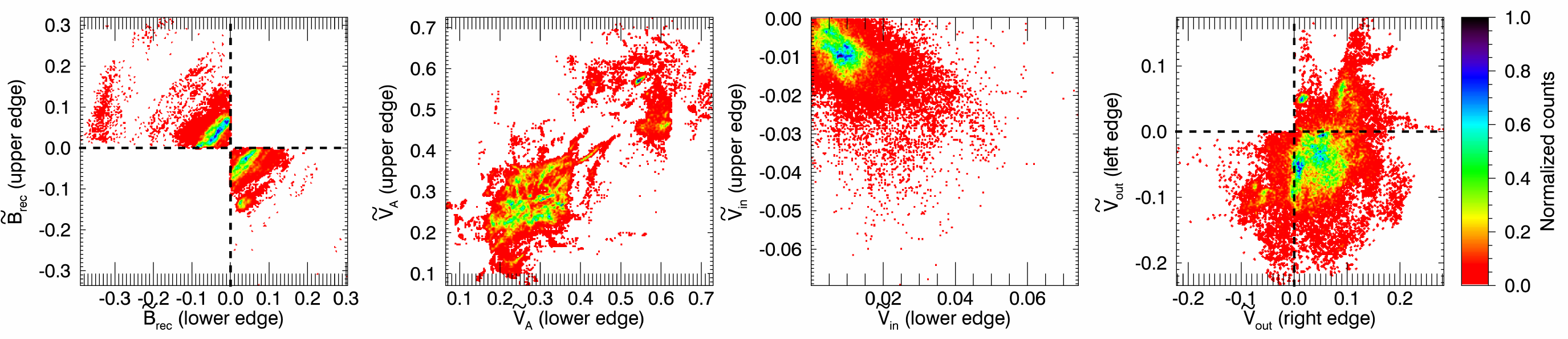}
  \caption{The diagrams show normalized $2D$ histograms (with $200$ bins in each direction) of the reconnected magnetic components (first diagram from the left to the right), the Alfv\'{e}n speed (second diagram), and the inflow (third diagram) and outflow (forth diagram) velocities (in units of $c$) at the upper and lower edges of the diffusion regions (in the reconnection reference frame, tilded quantities). The axes correspond to the values of each variable at the upper and lower edges, and the colorbar corresponds to the normalized counts.}
  \label{fig:edges}
\end{figure*}

Figure \ref{fig:edges} shows, from the left to the right, the normalized $2D$ histograms (with $200$ bins in each direction, and considering all events during the system's evolution) of the reconnected magnetic field components, the Alfv\'{e}n speed, the inflow and outflow velocities (all in units of $c$) at the upper and lower edges of the diffusion region (see Figure \ref{fig:scheme_jet} for the reference model m240ep0.5\footnote{For this model, the edges have been defined as the position where the current density value decreases to half of its maximum value at the center of the diffusion region ($|\boldsymbol{J}_{edge}| = 0.5|\boldsymbol{J}_{max}|$; see Section \ref{sec:algorithm} and Table \ref{tab:parameters}).}). 
The axes correspond to the values of each variable at the upper and lower edges, and the color bars correspond to the normalized counts. The shapes of the first and third histograms are due to selection effects, because we have constrained the sample to characterize the opposite magnetic field and inflow velocity components on each side of the current sheets in the reconnection events (see section \ref{sec:algorithm}). 
In particular, the signs of the inflow velocity at the upper edge must be negative, and positive at the lower edge (as we see in the third diagram) due to the direction of the eigenvectors of the local coordinate system obtained from the Hessian matrix. Also, considering this local system of coordinates, in the evaluation of outflow velocity in the forth histogram,
 we  removed from the sample, sites with positive velocity at the left edge and negative at the right edge. This because in such cases  the plasma is moving back into the diffusion region instead of going out. Nevertheless,  we have kept in the histogram the cases with outflow velocities with the same sign both on the left and right edges since at least one of the edges shows an outflow profile. The second diagram from left shows that the Alfv\'{e}n speed values are between $0.07\,c$ and $0.71\,c$, with more dominant cases in a range of $0.17\,c$ to $0.38 \,c$. The third and forth diagrams indicate mildly relativistic velocities with the highest inflow velocity $\tilde{V}_{in}\sim 0.075\,c$, and the highest outflow velocity $\tilde{V}_{out} \sim 0.28\,c$ (for all sites, the outflow velocity is always higher than the inflow velocity, as one might expect). All the diagrams of Figure \ref{fig:edges} show high asymmetric events at the diffusion region's edges that should be taken into account for the evaluation of the reconnection rate \citep[see, e.g.,][]{cassak_etal_11}, as we will see in the next section.

\begin{figure}
\centering
  \includegraphics[scale=0.38]{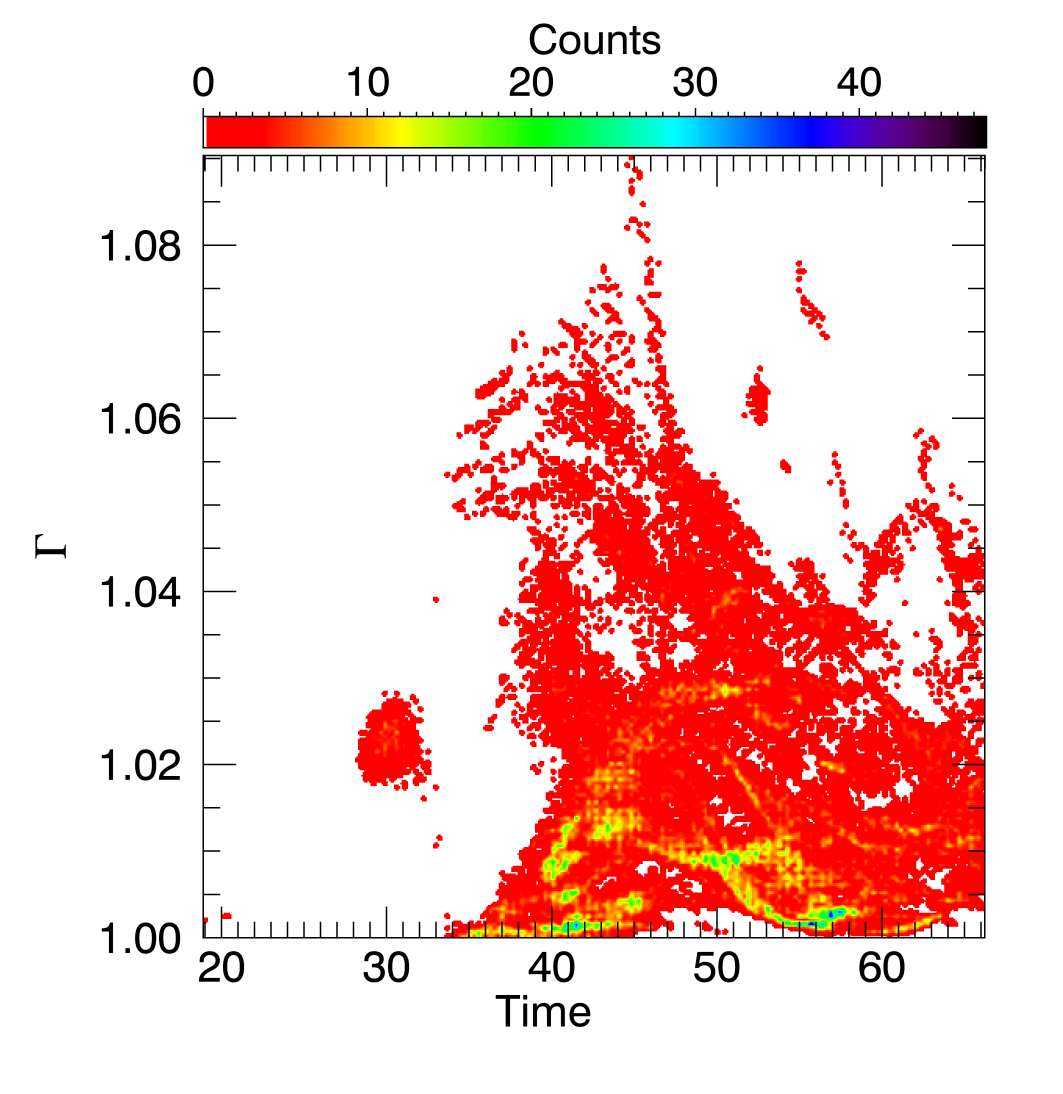}
  \caption{Time-distribution of the Lorentz factor ($\Gamma$) of the diffusion regions with respect to the jet frame, for the model m240ep0.5. The histogram corresponds to Lorentz factors obtained during the entire evolution of the simulation (with 200 bins in each direction).}
  \label{fig:lorentz}
\end{figure}

Finally, we note that, as we can see in the second diagram of Figure \ref{fig:edges}, the high values of the Alfv\'{e}n speed at the reconnection frame justify the relativistic correction for this quantity. However, the diffusion regions velocities are mildly relativistic with respect to the jet frame. Figure \ref{fig:lorentz} shows that the time-distribution of the Lorentz factor ($\Gamma$) in the diffusion regions is not higher than $1.09$. Thus although we have used the Lorentz transformation to obtain the velocity profiles around the diffusion region in the reconnection reference frame, such correction will not change significantly with respect to the Galilean transformation.

\subsection{Magnetic Reconnection Rate}
\label{sec:vrec}

We evaluate the reconnection rate as the ratio between the inflow velocity and the Alfv\'{e}n speed ($\tilde{V}_{rec}=\tilde{V}_{in}/\tilde{V}_{A}$) at the upper and lower edges of the diffusion region (in the reconnection reference frame). Furthermore, since the reconnection sites in this turbulent system are highly asymmetric events, as we saw in the previous section, a slight correction has been performed where a weighted arithmetic mean is used to obtain a single value for each site, so that
\begin{equation}
\langle \tilde{V}_{rec} \rangle =
\frac{[\tilde{h}_r|\tilde{V}_{e_1}|/\tilde{V}_{A}]_{lower}+[\tilde{h}_r|\tilde{V}_{e_1}|/\tilde{V}_{A}]_{upper}}{[\tilde{h}_r]_{lower}+[\tilde{h}_r]_{upper}}~,    
\end{equation}
where $\tilde{h}_r$ is distance of the (lower or upper) edges to the center of the diffusion region (see last sketch of Figure \ref{fig:scheme_jet}), and $|\tilde{V}_{e_1}|$ the absolute inflow velocity in the $\tilde{e}_1$ direction at the reconnection edges.   

\begin{figure}
  \includegraphics[scale=0.27]{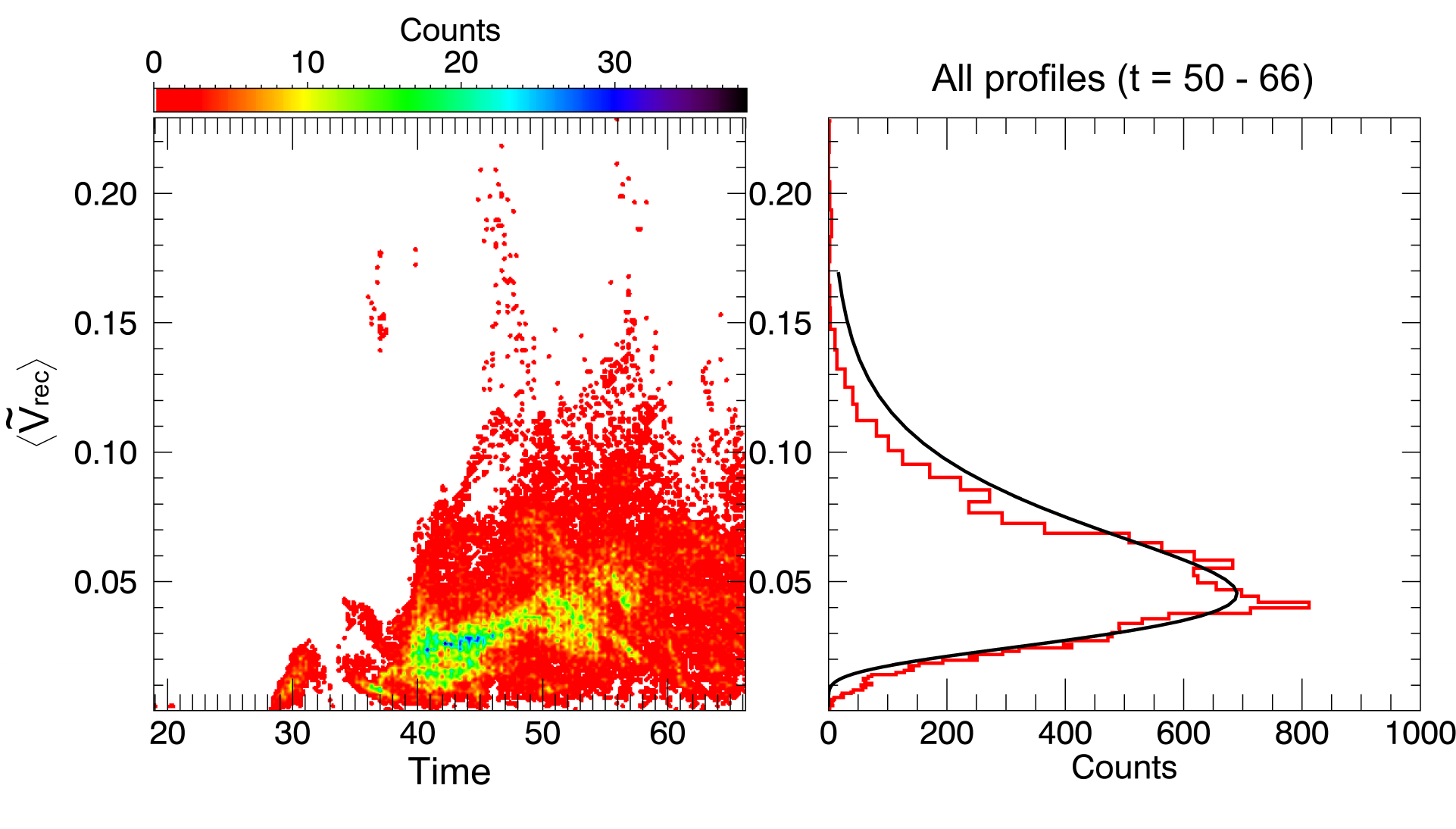}
  \includegraphics[scale=0.27]{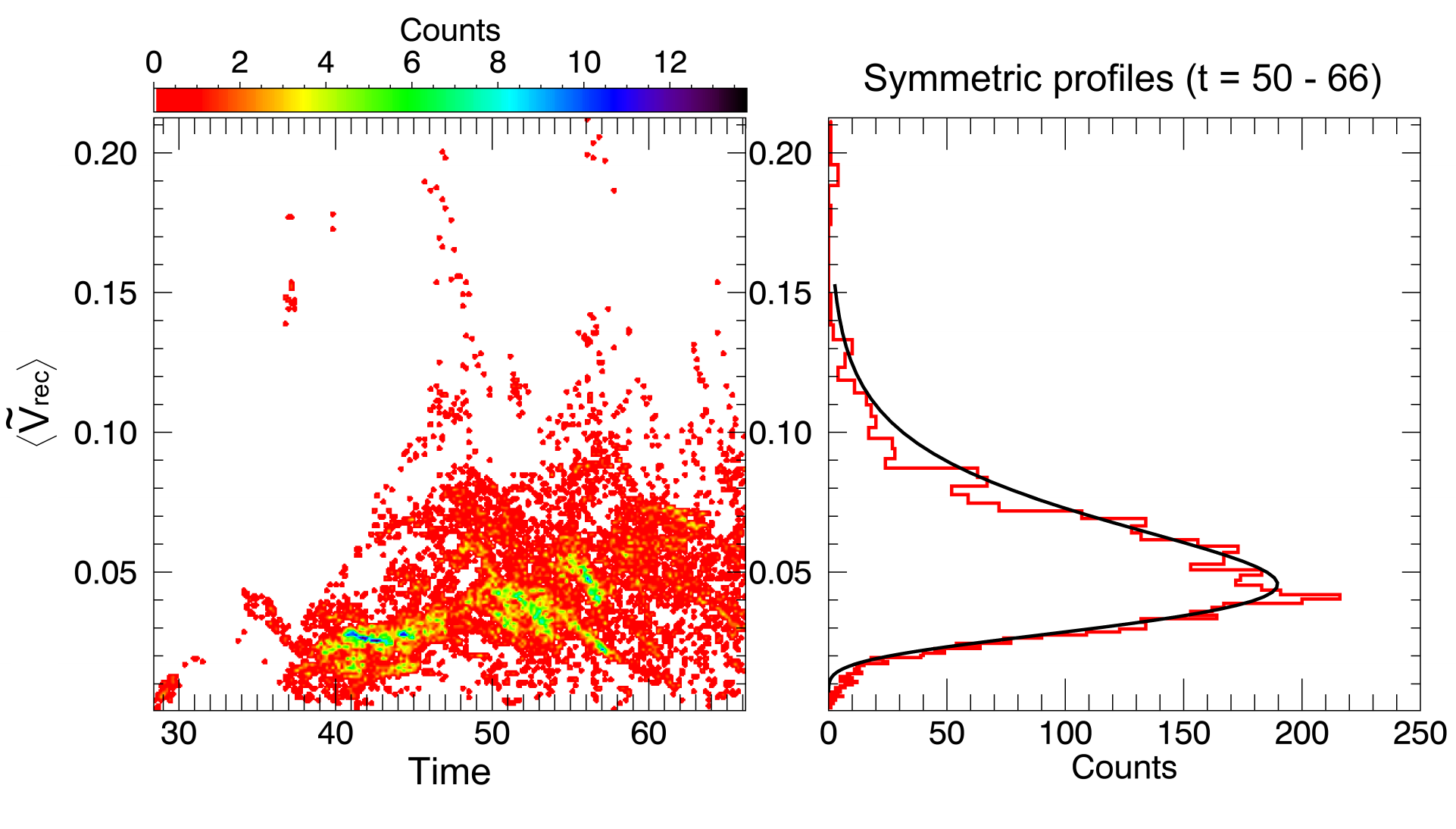}
  \caption{Histograms of $\langle \tilde{V}_{rec} \rangle$ for the model m240ep0.5. The left diagrams correspond to time-distributions obtained during the entire evolution of the simulation (with $200$ bins in each direction), and the right diagrams correspond to $1D$-distributions (red lines) with a log-normal fit (black lines) obtained between the snapshots $50$ and $66$ (with $140$ bins). The top distributions correspond to the whole sample, and the bottom ones correspond to a constrained sample considering only the most symmetric profiles of the velocity and magnetic fields at the edge of the diffusion regions.}
  \label{fig:vrec_hist}
\end{figure}

Figure \ref{fig:vrec_hist} shows different histograms of $\langle \tilde{V}_{rec} \rangle$ for the model m240ep0.5. The left diagrams correspond to time-distributions obtained during the entire evolution of the simulation\footnote{We highlight that the time interval studied in the present work corresponds to the entire dynamical time of the kink instability (from the nonlinear growth to the saturation and jet disruption).} (with $200$ bins in each direction), and the right diagrams correspond to $1D$-distributions obtained between $t=50$ and $66$ (with $140$ bins), when the system is already in a quasi-steady turbulent regime \citepalias[see also][]{medina_torrejon_etal2020}. 
The left top diagram shows sporadic and slow reconnection events ($\langle \tilde{V}_{rec} \rangle < 0.01$) around $t=20$. After $t \sim 35$, the number of events and the value of $\langle \tilde{V}_{rec} \rangle$ increase as the system reaches the saturation of the exponential growth of the CDK instability \citepalias[see section 3.1 in][]{medina_torrejon_etal2020}. Before that, we note that there is a lack of events, particularly between $t = 33$ and $34$. This because at these times the CDK is still growing and the magnetic field is essentially getting distorted, but with almost no reconnection.
More suspicious are those events found around $t=20$, which are very slow and possibly are not real reconnection layers. After $t \sim 35$, 
$\langle \tilde{V}_{rec} \rangle$ 
increases, having the fastest rate of $\langle \tilde{V}_{rec} \rangle \sim 0.23$ at $t=56$. 

The right top histogram of Figure \ref{fig:vrec_hist} shows that the distribution of $\langle \tilde{V}_{rec} \rangle$ does not resemble a normal distribution, showing a long tail on the side of the fastest rates \citep[similar to the results obtained by][for non-relativistic accretion disk systems]{kadowaki_etal_18}. This skewed feature is characteristic of a log-normal distribution, and to test this hypothesis we have performed fits (black lines in the right histograms) and the results are presented in Table \ref{tab:fitvrec}. 
We have also compared the results of the fit with the four statistical moments of the sample (i.e., mean, variance, skewness, and kurtosis moments). In both histograms, we obtained for the skewness and kurtosis non-zero (positive) values, characteristic of a skewed distribution with a peaked shape near the mean, as we expected. Furthermore, we applied a reduced chi-square statistic to evaluate the quality of the fit, obtaining the poor value of $\chi^{2}_{red} \sim 17$ (see Table \ref{tab:fitvrec}). Based on this result, we constrained the sample in order to obtain the reconnection rate only for the most symmetric profiles (bottom histograms of Figure \ref{fig:vrec_hist})\footnote{We note that for the symmetric sample, the magnetic and velocity magnitudes at one edge of the diffusion region will not be two times larger than the values in the opposite edge \citep[as in][]{kadowaki_etal_18}.}. This constraint reduces to one-forth the sample size, but also implies in a reduced chi-square to $\chi^{2}_{red} \sim 2.8$ (see Table \ref{tab:fitvrec}). Despite this difference, we obtained from the fit of the constrained sample an average reconnection rate $\langle \tilde{V}_{rec} \rangle_f$ of the order of $0.050 \pm 0.021$, which does not differ from the value of the original sample ($\langle \tilde{V}_{rec} \rangle_f = 0.051 \pm 0.026$, see Table \ref{tab:fitvrec}). Furthermore, this value is compatible with that found in \citet{singh_etal_16}.

\input{table01.tex}

\begin{figure}
\centering
  \includegraphics[scale=0.38]{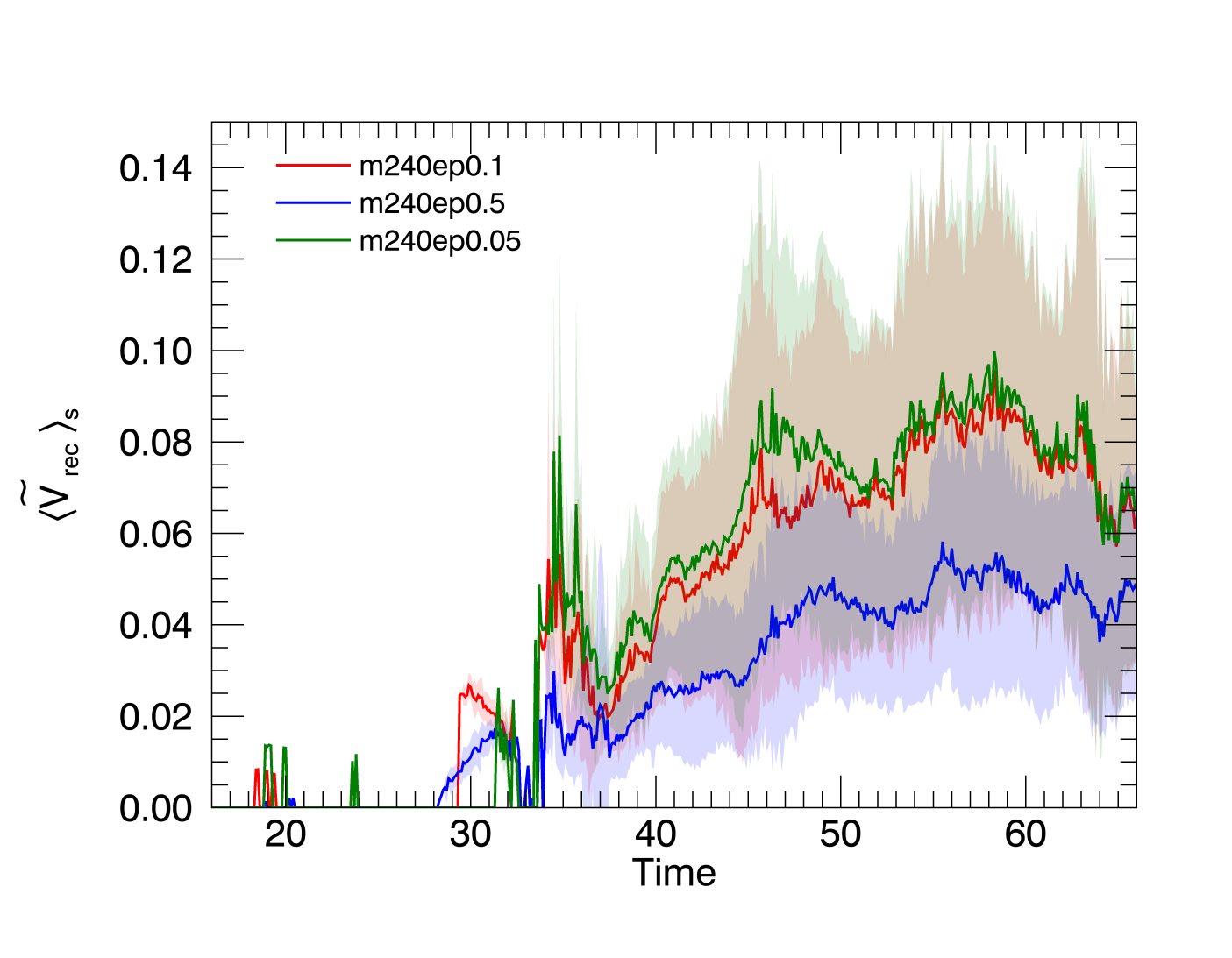}
  \includegraphics[scale=0.38]{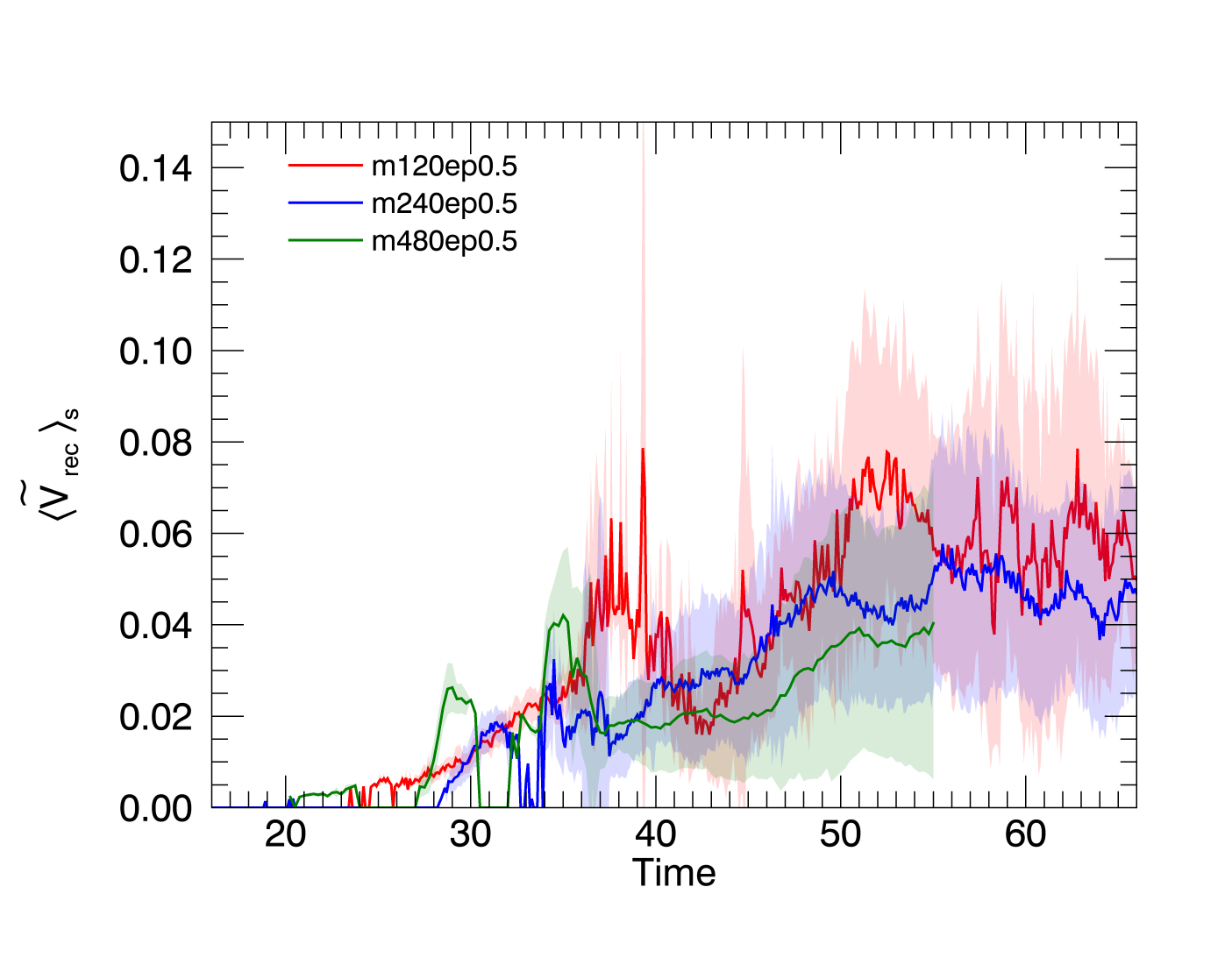}
  \caption{The top diagram shows the time evolution of the average magnetic reconnection rate $\langle \tilde{V}_{rec} \rangle_s$ (taken from all identified sites in each snapshot) considering different criteria for the position of the edges of the diffusion region ($0.5$, $0.1$, and $0.05 |\boldsymbol{J}_{max}|$), whereas the bottom diagram compares the time evolution of $\langle \tilde{V}_{rec} \rangle_s$ for three different resolutions of the jet simulation ($120^3$, $240^3$, and $480^3$ cells). Colored shades correspond to standard deviations of each model.}
  \label{fig:vrec_comp}
\end{figure}

We have also obtained the time evolution of the average magnetic reconnection rate, $\langle \tilde{V}_{rec} \rangle_s$ (taken from all identified sites in each snapshot), 
considering different criteria for the position of the edges of the diffusion region ($0.5$, $0.1$, and $0.05 |\boldsymbol{J}_{max}|$; see Section \ref{sec:algorithm}). This is shown in the top diagram of Figure \ref{fig:vrec_comp}, where each line corresponds to the models m240ep0.5 (blue line), m240ep0.1 (red line), and m240ep0.05 (green line; see Table \ref{tab:parameters}). In general, the evolution of $\langle \tilde{V}_{rec} \rangle_s$ does not change significantly between $t=16$ and $34$, when the CDK instability grows exponentially \citepalias[see][Figure 2]{medina_torrejon_etal2020}. All the models show few and slow events with a peak  around $t=32$, and a gap between $t=33$ and $34$. After this time, the models m240ep0.1 and m240ep0.05 show convergence, but with $\langle \tilde{V}_{rec} \rangle_s$ values larger than those obtained for the reference model m240ep0.5 (at the limit of 1$\sigma$ uncertainty, indicated by the colored shades in Figure \ref{fig:vrec_comp}), particularly after the CDK instability has achieved the saturation and  quasi-steady-state turbulence is settled in the system, beyond $t=40$. The reconnection regions in the models m240ep0.1 and m240ep0.05 are larger than in model m240ep0.5 since the asymptotic magnetic field and the velocity have been measured at locations where the current density decays to $0.1$ and $0.05$ of the maximum, respectively. Despite the convergence of the models m240ep0.1 and m240ep0.05, we adopted for our reference model the conservative criterion of $0.5|\boldsymbol{J}_{max}|$ (model m240ep0.5), also used in previous works \citep[see, e.g.,][]{kowal_etal_09,zhdankin_etal_13,kadowaki_etal_18}. In fact, we do not expect a convergence of $\langle \tilde{V}_{rec} \rangle_s$ in such comparisons since we can extrapolate the size of a single identified region.   

The bottom diagram of Figure \ref{fig:vrec_comp} compares the time evolution of $\langle \tilde{V}_{rec} \rangle_s$ obtained for the reference jet model with resolution of $240^3$ cells (model m240ep0.5, blue line), and with two lower and higher resolution models, i.e., $120^3$ (model m120ep0.5, red line)  and $480^3$ (model m480ep0.5, green line) cells, respectively.
The lowest resolution model (m120ep0.5) has the largest differences, with a smooth increase of $\langle \tilde{V}_{rec} \rangle_s$ (with no gaps) between $t=23$ and $36$. 
After this time, the model m120ep0.5 achieves a peak around $t \sim 40$, i.e., when the CDK instability reaches the plateau, followed by high variability, but with average  fast rates over the quasi-steady-state turbulent regime, as in the models m240ep0.5 and m480ep0.5. The highest resolution model (m480ep0.5)\footnote{The model m480ep0.5 is numerically expensive and unstable, thus we have evolved it until t=$55$.} shows the smallest variability, and a gap between $t=31$ and $32$ (earlier than that one obtained in the reference model m240ep0.5). Furthermore, it converges approximately to the model m240ep0.5 after $t=40$ (within $1\sigma$ uncertainty). The latter results are in agreement with the earlier works of \citet{mizuno_etal_09,mizuno_etal_12,mizuno_etal_14}, where the numerical convergence was carefully tested and found between the intermediate and high-resolution models, but not for the lowest one. 
The absence of significant differences between the models of the bottom diagram are also compatible with the fact that fast magnetic reconnection driven by turbulence is  independent of the numerical resistivity \citep[as obtained by][]{kowal_etal_09, kowal_etal_2012,kadowaki_etal_18}. This is also compatible with the
turbulence-induced fast reconnection theory of \citet[][]{lazarian_vishiniac_99}, that predicts that the turbulence speeds up the reconnection independently of the Ohmic resistivity of the environment. 

\begin{figure}
\centering
  \includegraphics[scale=0.38]{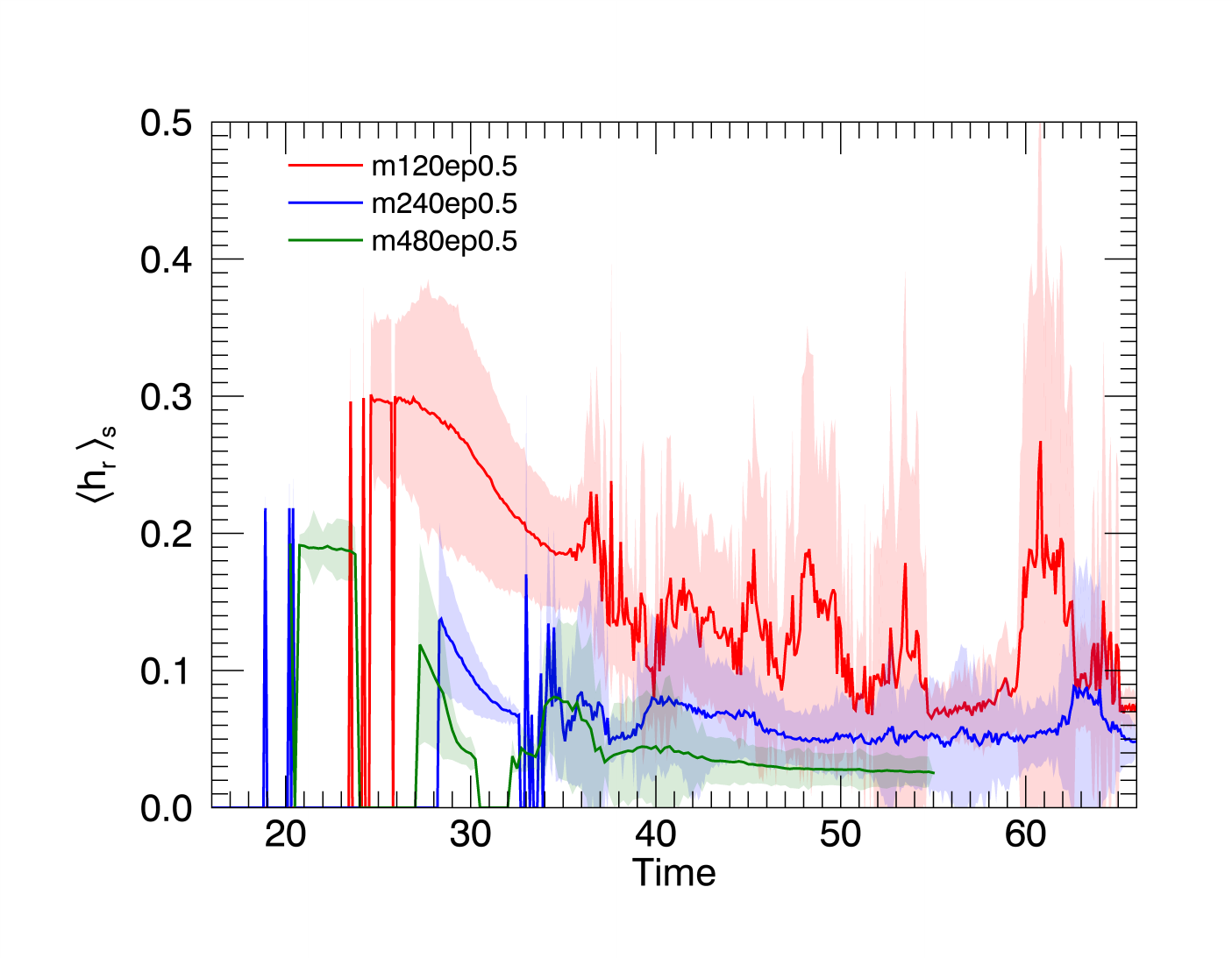}
  \includegraphics[scale=0.38]{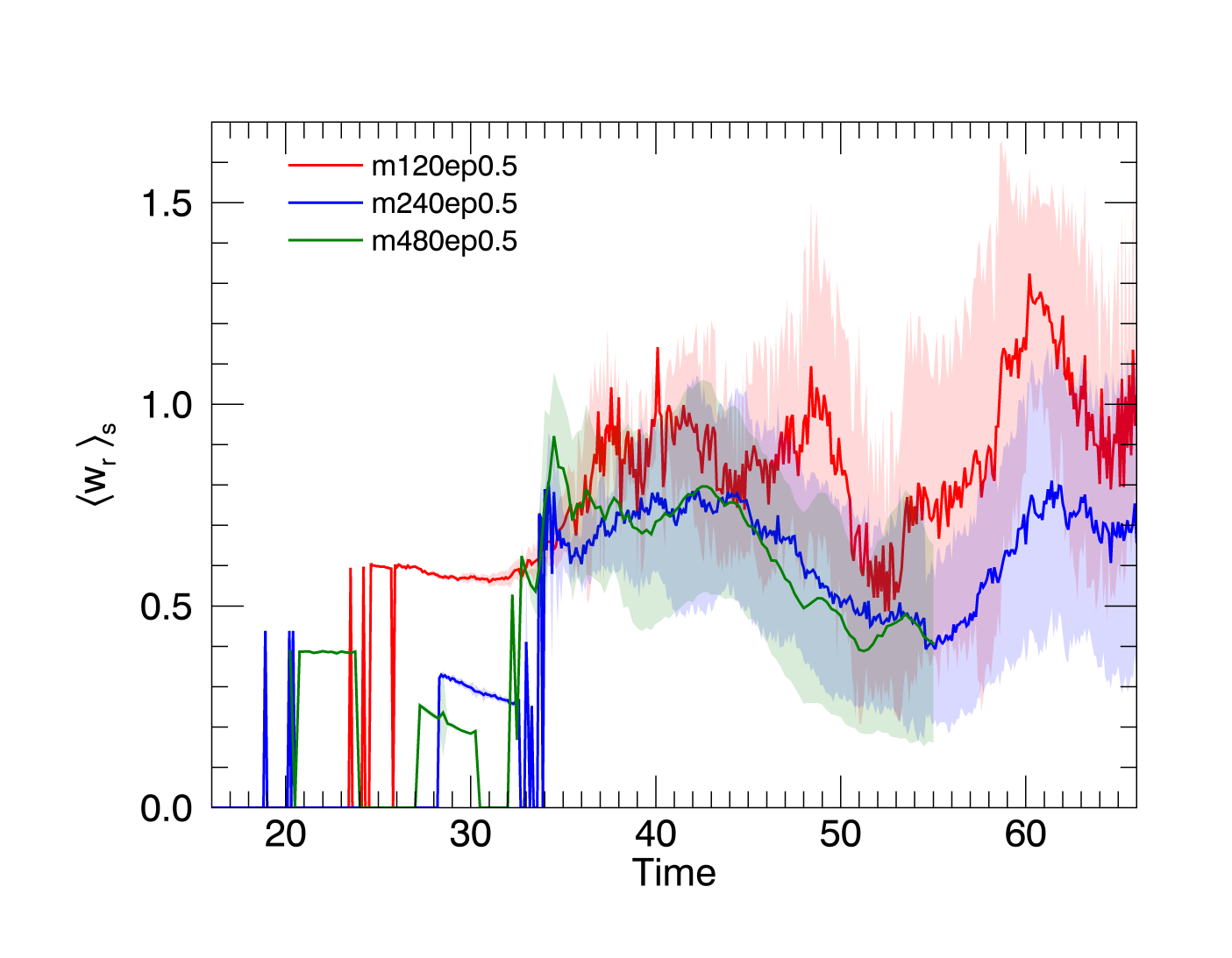}
  \caption{Time evolution of the average height (top diagram) and width (bottom diagram) measured in the magnetic reconnection sites at jet frame. As in the bottom diagram of Figure \ref{fig:vrec_comp}, the lines correspond to the three different resolutions of the jet simulation ($120^3$, $240^3$, and $480^3$). Colored shades correspond to standard deviations of each model.}
  \label{fig:hw_comp}
\end{figure}

In order to verify the resolution effects in the estimates of the width and height of the magnetic reconnection sites, we have also evaluated the time-evolution of the average values of these quantities. Figure \ref{fig:hw_comp} compares these quantities, where we included the standard deviations of the average values for each model (colored shades; as in the diagrams of Figure \ref{fig:vrec_comp}). As we might expect, the intermediate (m240ep0.5) and high-resolution (m480ep0.5) models also show convergence within  $1\sigma$ uncertainty after t=40, which indicates that the $\langle \tilde{V}_{rec} \rangle_s$ behavior shown in the top diagram of Figure \ref{fig:vrec_comp} for different edge criteria will not change significantly in the case of the higher resolution simulations.

Finally, as mentioned before, the log-normal distribution can be associated with the turbulence. Lately, authors have been using the four statistical moments to characterize compressible MHD turbulence from simulations, evaluating its relation with different sonic and Alfv\'{e}nic Mach numbers, especially for the interstellar medium \citep[see, e.g.,][and references therein]{kowal_etal_07,burkhart_etal_09,burkhart_etal_17, BarretoMota_etal_21}. In these works, the density distribution of molecular clouds in the presence of turbulence naturally converges to a log-normal profile. It is interesting to note that the $\langle \tilde{V}_{rec} \rangle$ distribution also follows this profile (Figure \ref{fig:vrec_hist}), at the same time that the averaged reconnection rate becomes $fast$ (Figure \ref{fig:vrec_comp}), i.e., in the quasi-steady-state turbulent regime (as we can see in the last diagram of Figure \ref{fig:jetevol}). 
Such behavior also occurs in accretion disks where turbulence induced by  magneto-rotational and Parker-Rayleigh-Taylor instabilities drive fast reconnection \citep{kadowaki_etal_18}. Likewise, the results obtained in this section indicate that the turbulence is the process driving fast reconnection events \citep[as predicted by ][]{lazarian_vishiniac_99} in the relativistic jet, under the action of current-driven kink instability. Other systems where turbulence is excited by distinct physical mechanisms such as Kelvin-Helmholtz and  Weibel instabilities  also experience fast reconnection \citep[e.g.][]{kowal_etal_2020, nishikawa_etal_2020}.

\subsection{Turbulence Power Spectra}

In the previous section, we showed that the $1D$-histograms of $\langle \tilde{V}_{rec} \rangle$ resemble a log-normal distribution when evaluated at times between $t\simeq50$ and $66$ (Figure \ref{fig:vrec_hist}) which can be associated with turbulence \citep[see, e.g.,][]{burkhart_etal_09,burkhart_etal_17, BarretoMota_etal_21}. In order to further quantify the presence of  turbulence, we have performed a Fourier analysis and evaluated the three-dimensional power spectra of the magnetic and kinetic energy densities. These have been taken from the average of spherical shells between $k$ and $k + dk$ (where $k=\sqrt{k_x^2+k_y^2+k_z^2}$) in the Fourier space, and are given by \citep[see, e.g.,][]{simon_etal_09}:

\begin{equation}
    {1\over2}|\boldsymbol{B}(\boldsymbol{k})|^2={1\over2}[|B_x(\boldsymbol{k})|^2+|B_y(\boldsymbol{k})|^2+|B_z(\boldsymbol{k})|^2]~, \textrm{and}
\end{equation}
\begin{equation}
    {1\over2}|\sqrt{\rho}\boldsymbol{v}(\boldsymbol{k})|^2={1\over2}[|\sqrt{\rho}v_x(\boldsymbol{k})|^2+|\sqrt{\rho}v_y(\boldsymbol{k})|^2+|\sqrt{\rho}v_z(\boldsymbol{k})|^2]~.
\end{equation}   

\begin{figure}
\centering
  \includegraphics[scale=0.52]{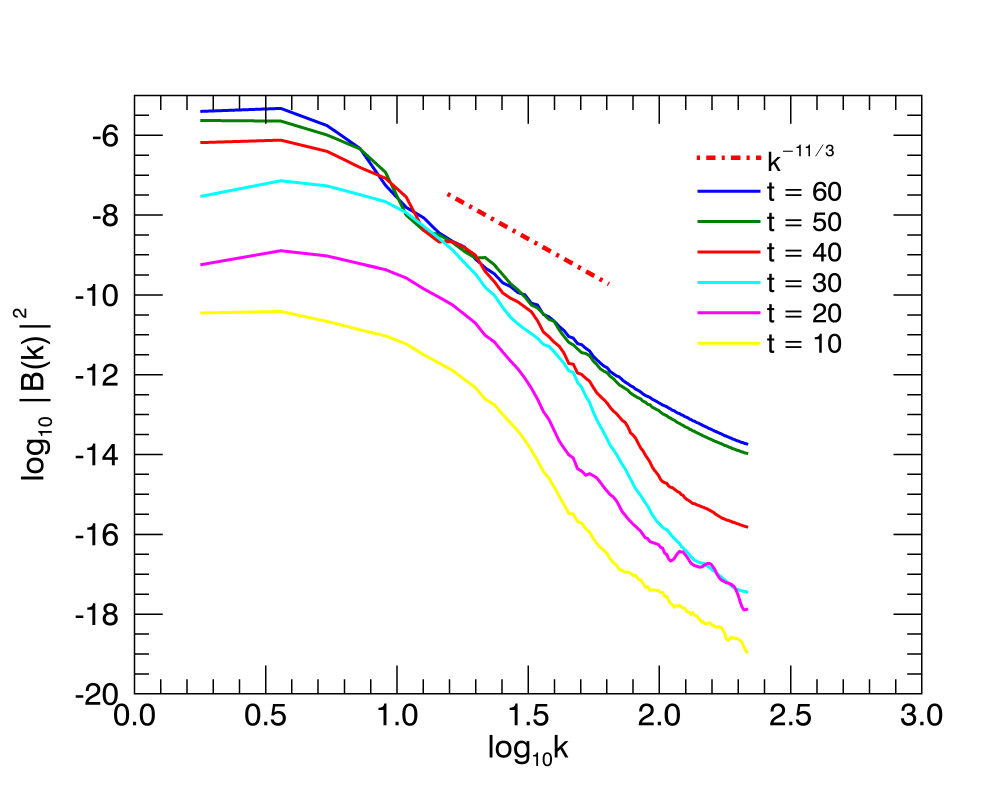}
  \includegraphics[scale=0.52]{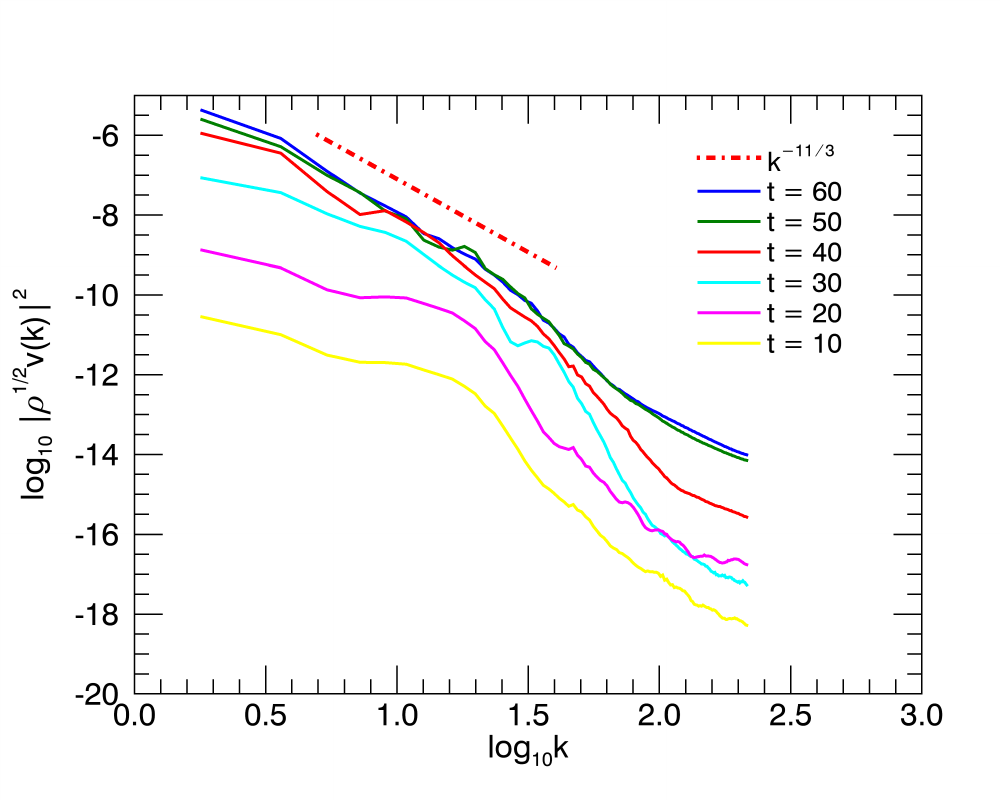}
  \caption{Power spectra of the magnetic (top diagram) and kinetic (bottom diagram) energy densities for different times ($t=10$, $20$, $30$, $40$, $50$ and $60$ represented by colored continuous lines). The red dashed line corresponds to the $3D$-Kolmogorov spectrum $k^{-11/3}$.}
  \label{fig:PS_turb}
\end{figure}

Figure \ref{fig:PS_turb} shows the power spectra for the magnetic (top diagram) and kinetic (bottom diagram) energy densities for different times (from $t=10$ to $60$; colored continuous lines). A $3D$-Kolmogorov spectrum ($\propto k^{-11/3}$; red dashed line) was included for comparison. The diagrams show inertial ranges with slopes ($k^\nu$) between $-3.7$ and $-3.6$ for $|\sqrt{\rho}\boldsymbol{v}(\boldsymbol{k})|^2$ (from $\log_{10} k =0.7$ to $\log_{10} k=1.4$), and $-5.1 < \nu < -3.3$ for $|\boldsymbol{B}(\boldsymbol{k})|^2$ ($1.1 < \log_{10} k < 1.4$), both in agreement with a Kolmogorov-like spectrum after $t\simeq50$ (green and blue lines), indicating a turbulent energy cascade.  The magnetic energy spectrum shows steeper slopes, probably due to the strong guiding magnetic  field which is still present at later times \citep[see, e.g.,][]{kowal_etal_07}. We also note  that at the time that the CDKI achieves the saturation ($t=40$), a nearly Kolmogorov-like shape is already present in the inertial range of the spectrum, though it is not yet as well defined as in later times, thus  indicating that the turbulence is still not fully developed at this time.
These results strengthen those shown in section \ref{sec:vrec}
and  Figure \ref{fig:vrec_hist} at the same time-interval. Furthermore, We also notice that $\langle \tilde{V}_{rec} \rangle_s$ achieves a quasi-steady-state behavior with the fastest rates between $t=50$ and $66$ (see Figure \ref{fig:vrec_comp}), demonstrating the correlation between turbulence and fast magnetic reconnection events.    

\section{Magnetic Reconnection: An application to Mrk 421}
\label{sec:application}

In this section, as an example, we consider an application of our general results obtained for magnetized relativistic jets to the blazar source Mrk 421. Based on the study of \citet[][]{kushwaha_etal_17}, who analyzed the light curve of this source and its variability features in the high energy band between $0.1-300$ GeV, we try here to reproduce it, assuming that this emission pattern is associated to fast magnetic reconnection events.

As we have stressed in Section \ref{sec:intro}, regions of fast reconnection are sites of efficient stochastic particle acceleration \cite[e.g.,][]{dalpino_lazarian_2005,kowal_etal_2012,delvalle_etal_16}. 
Our simulations do not allow us to evaluate the exact fraction of released magnetic energy by reconnection that goes into particle acceleration. Nevertheless, in the companion paper \citetalias{medina_torrejon_etal2020}, we have injected  test particles  within the same  SRMHD background jet model studied here, and found that particles with an initial energy $10^{-4} m_pc^2$ are efficiently accelerated in the turbulent fast reconnection current sheets of the system up to very high energies $\sim 10 ^{7}$ - $10 ^{9}$ $m_pc^2$, after several hundred hours (for background magnetic fields with $\sim 0.1$ - $10$  $G$), where $m_p$ is proton mass. These results indicate that the fast magnetic reconnection events not only accelerate the relativistic particles, but also could account for the observed very-high-energy variable emission that these particles may produce in the magnetically dominated regions of blazars.

With this in mind, we here make the  
assumption that the variability pattern of the reconnection events that we computed can be directly correlated with the variable emission in the high-energy band of sources like the one above, which is expressed in its light-curve.

We start by considering that in the dissipation zone of the blazar, roughly half of the magnetic reconnection power is converted into kinetic power accelerating efficiently the particles to very-high-energies \citepalias[as found in][]{medina_torrejon_etal2020}, which will then be able to produce high-energy photons.  This assumption is compatible with the fact that about 50\% of the energy released by reconnection goes into particle acceleration
\citep[e.g.,][]{yamada_etal2016} 
and thus can be used to constrain the radiative power as well \citep[see also, e.g.,][]{christie_etal_19}.

We should also emphasize that it is out of the scope of this  work to discuss the specific radiative non-thermal  processes that may lead to photon production by the accelerated particles. This would require a full study involving radiative transfer in the jet background and considerations on  the  leptonic and/or hadronic composition  of the source. Our purpose here is only to benchmark the robustness of our reconnection search method and 
verify to what extent the reconnection events can be connected with the variability  and emission patterns of these relativistic sources. 
In forthcoming work, we plan to incorporate radiative transfer effects in our SRMHD code in order to perform a complete reconstruction of observed spectral energy distributions of blazar jets, in a similar way as performed, for instance, in \citet[][]{rodriguezramires_etal_18}.

Basically, we evaluate here the magnetic power $\tilde{L}_{B}(t_s)$ released from the reconnection sites (in the reconnection frame) in a snapshot $t_s$, as follows:

\begin{equation}
\label{eq:power}
   \tilde{L}_{B}(t_s) = \sum^{N(t_s)}_{i=1}{\langle {\tilde{B}_{in}^2 \over 2} \tilde{V}_{in} \tilde{w}_{r}^2 \rangle}_{i},
\end{equation}
where $\tilde{B}_{in}$ is the reconnected magnetic component, $\tilde{V}_{in}$ is the inflow velocity, and $\tilde{w}_{r}^2$ is the area of the reconnection site at the edges of the diffusion region, measured in the reconnection frame (see the last sketch of Figure \ref{fig:scheme_jet}). 
We assume that the emission comes from  the entire  simulated  domain (see Figure \ref{fig:scheme_jet}). To obtain $\tilde{L}_{B}(t_s)$, we summed the contributions of all $N(t_s)$ sites identified by the algorithm in the snapshot $t_s$. Since the velocities of the reconnection sites are mildly relativistic at the jet frame, with  Lorentz factors of the order of unity ($\Gamma_r \sim 1$; see Figure \ref{fig:lorentz}), then the neglect of a length contraction from Lorentz transformation will not change significantly the results. Therefore, we can estimate the area by the square of the width of the reconnection site measured in the jet frame ($\tilde{w}_{r} \sim w_{r}$). Likewise, we can also evaluate the jet luminosity in the dissipation zone as $L_j(t_s) \sim \eta\tilde{L}_{B}(t_s)$, where $\eta=0.5$ is the  efficiency of conversion of the magnetic power into jet luminosity, as discussed above.

We consider that the simulated jet is moving relative to an observer's reference frame (primed quantities) with a bulk Lorentz factor $\Gamma^{\prime}_j$, and an angle $\theta^{\prime}_j$ to the line of sight (see the first sketch of Figure \ref{fig:scheme_jet}). This assumption is reasonable as long as the reconnection sites are mildly relativistic at the jet frame. Therefore, we have also considered the observed luminosity $L^{\prime}_j(t_s)=\delta^{\prime4} L_j(t_s)$ related to the comoving luminosity $L_j(t_s)$ (assumed to be isotropic at the jet frame), where $\delta^{\prime} = [\Gamma^{\prime}_j(1-\beta^{\prime}_j cos\theta^{\prime}_j)]^{-1}$ is the relativistic Doppler factor, and $\beta^{\prime}_j$ is the jet bulk velocity. We have also evaluated the time intervals measured in the observer frame as $t^{\prime}=\delta^{\prime-1} t$, where $t$ refers to the jet frame.

\subsection{Physical units and synthetic light curves}
\label{sec:phunit_lightcurve}

As stressed, for the jet simulation we used the 3D general relativistic MHD code \texttt{RAISHIN} \citep{mizuno_etal_06} that works with scale-free (code) units. For the comparison with the observations, $L_j(t_s)$ must be converted into physical units. To this aim, we have specified three physical constant units, velocity, time, and magnetic field ($v_{0}, t_{0}, B_{0}$), at the jet frame (corresponding to the simulation domain). Other important units are obtained from the previous ones, such as, the length unit  ($r_{0}=v_{0} t_{0}$), the  density unit ($\rho_{0}=B^{2}_{0}/(4\pi v^{2}_{0})$), and the power unit  ($L_{0}=B^{2}_{0} t^{2}_{0} v^{3}_{0}/4\pi$), all in cgs units.   

\begin{figure}
  \centering
  \includegraphics[scale=0.3]{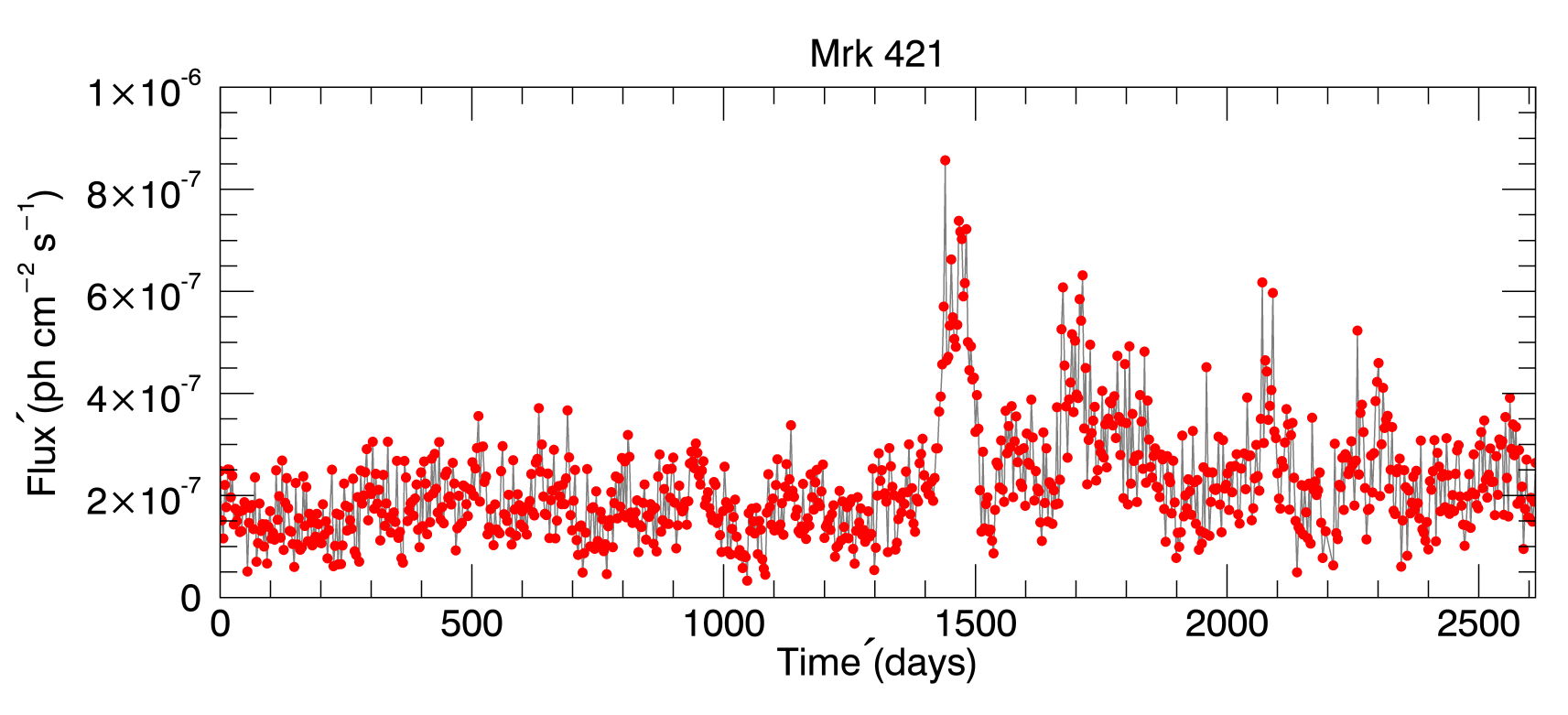}
  \includegraphics[scale=0.3]{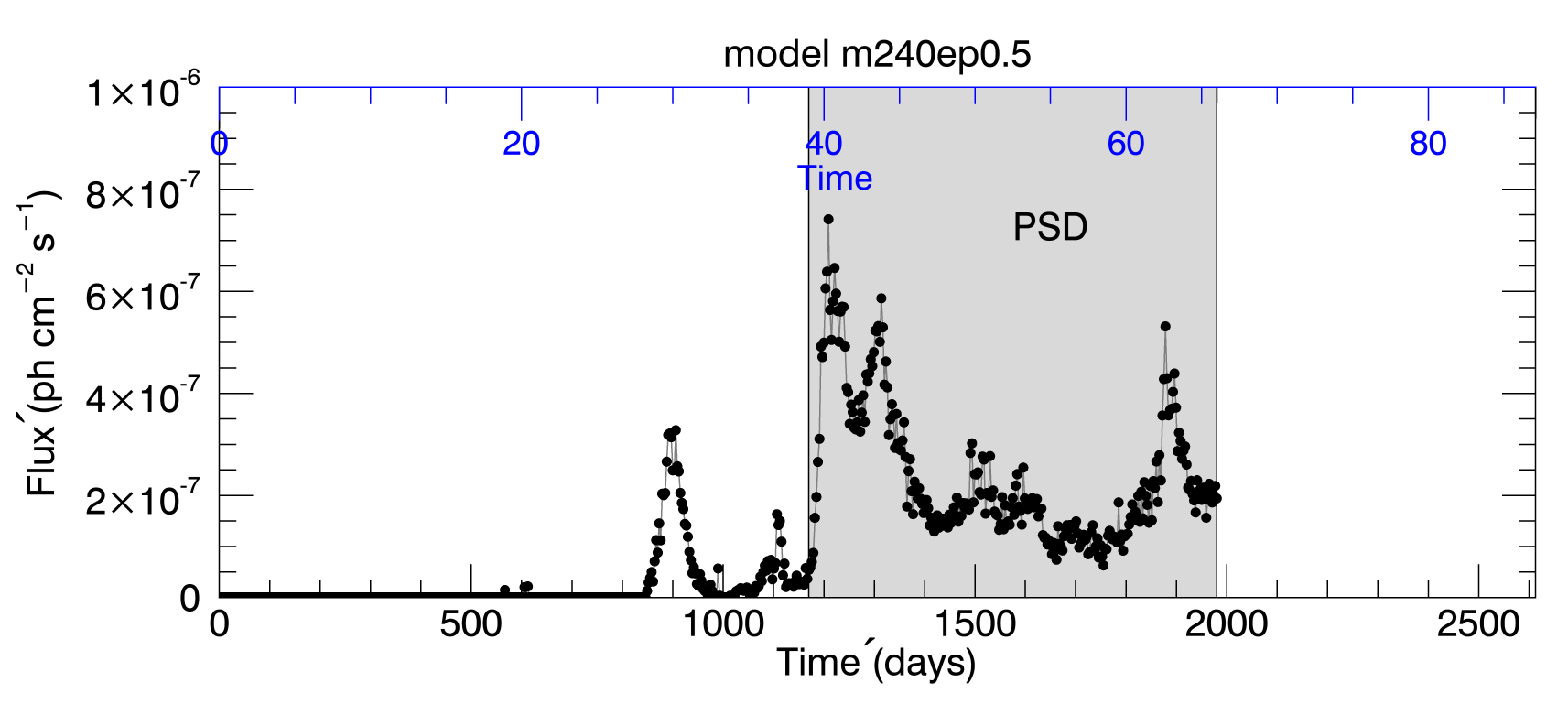}
  \caption{The top diagram shows the light curve of the AGN Mrk~421 \citep[in $0.1-300$ GeV band, obtained from][]{kushwaha_etal_17}. The bottom diagram shows the best synthetic light curve evaluated from our jet model m240ep0.5, where the time is expressed in cgs units at the observer's frame (bottom black horizontal axis with the primed symbol), and in code units at the jet frame (top blue horizontal axis). The shaded gray region corresponds to the data used for power spectral analysis (see more details in the text).}
  \label{fig:light_curves}
\end{figure}

We have employed the light speed as the  velocity unit ($v_{0}=c$) since the \texttt{RAISHIN} code already works with this normalization. Then, taking the light curve of BL Lac Mrk~421 \citep[see the top diagram of Figure \ref{fig:light_curves}; see also][]{kushwaha_etal_17} we have defined two other physical units. Assuming that the time step between each snapshot of the simulation is equal to the minimum time variability observed from Mrk 421 at the same frame, the time unit is thus given by:
\begin{equation}
\label{eq:physical_time}
t_{0} ={\delta^{\prime} \Delta t^{\prime}_{obs}[cgs]\over{\Delta t[c.u.]}}~,
\end{equation}
where $\Delta t$ (at the jet frame) corresponds to the minimum time variability from the simulations, in code units (c.u.), and $\Delta t^{\prime}_{obs}$ (at the observer's  frame) is the observed one for Mrk~421, in cgs units. We have chosen the magnetic field unit $B_{0}$ so that the maximum photon flux of Mrk 421 coincides with the one obtained from the simulations (in physical units). Once defined $B_{0}$, we have obtained the unit power $L_{0}$, and converted $L_j(t_s)$ to cgs units ($L[cgs] = L[c.u.]L_{0}$). 

As described previously, we have evaluated the luminosity in the observer's frame by the Doppler factor, and then estimated the synthetic photon flux (i.e., units of photons $\mathrm{cm}^{-2}\,\mathrm{s}^{-1}$) in the $0.1-300$ GeV energy range\footnote{We have converted the energy flux into photon flux using a power-law function $dN/dE \propto E^{-\alpha}$, where $\alpha \sim 1.78$ is the average photon index obtained from \citet{abdo_etal_11}. We have also assumed the distance of 133\,Mpc for Mrk~421 to evaluate the fluxes \citep[see, e.g.,][]{sbarufatti_falomo_05}.}. Table \ref{tab:punits} shows the physical units for each variable obtained from this analysis taking into account different values of Doppler factors. Additionally, we included the corresponding proton density for $\rho_{0}$.   

\input{table02.tex}

We used the jet model m240ep0.5 to build the synthetic light curve (see the bottom diagram of Figure \ref{fig:light_curves}) with a typical Doppler factor $\delta^\prime \sim 5$. This model shows an initial peak at $t \sim 30$ ($\sim 900$ days at the observer frame) followed by a phase of low activity between $t=32$ and $39$. After this time, the activity increases with three prominent peaks at $t\sim 40$, $44$ and $63$. This behavior is similar to the average reconnection rate profile (see Figure \ref{fig:vrec_comp}) described in Section \ref{sec:vrec}. At the first peak, the average reconnection rate is not high (the maximum $\langle \tilde{V}_{rec}\rangle_s$ value is around $0.02$ at reconnection frame), with a small number of events (see Figure \ref{fig:vrec_hist}), which indicates that each reconnection site identified by the algorithm releases considerable amounts of magnetic energy. Indeed, as we can see in the second diagram of Figure \ref{fig:jetevol}, these events occur near the central axis of the jet where the magnetic field is high. The low activity phase, in turn, is associated to the decreasing number of reconnection events (as we can see in Figure \ref{fig:vrec_hist}), with an intermediary peak at $t \sim 37$. After $t \sim 39$, the high activity seen in the synthetic light curve coincides with the nonlinear phase of the kink instability, where several reconnection sites are identified, with $\langle \tilde{V}_{rec} \rangle$ up to $0.02$. The model m240ep0.5 is interrupted at $t\sim 66$ due to complete disruption of the jet. Therefore, we believe that the high activity phase observed in the light curve of Mrk~421 can be correlated with the non-linear phase of the kink instability that drives turbulence and fast magnetic reconnection events \citepalias[see also][]{medina_torrejon_etal2020}.
Once the jet is locally disrupted, the activity is expected to decrease. If the kink instability rises again, the high activity can be resumed, starting another cycle. We will discuss in more detail the possible correlation between the time-variability of Mrk~421 and reconnection events in the next section. 

\subsection{Power Spectral Density Analysis}
\label{sec:psd}

It is important to highlight that 
the aim of the analysis presented in the previous section was not to match exactly the synthetic light curve to the observations of Mrk~421. This because in this work, we did not perform background jet simulations specifically for this source, as they were intended to be more general. 
Nevertheless, this analysis has shown that  we can already obtain reliable values, when converting the results of the simulations in physical units. The choice of Mrk~421 is based on the work of \citet{kushwaha_etal_17}, where a careful study of this source was performed from a light curve power spectral density (PSD) analysis. The authors obtained the statistical proprieties of the time variability of Mrk~421, and concluded that it is broadly consistent with magnetic reconnection events, as described in the jets-in-a-jet model \citep[see, e.g.,][]{giannios_etal_09,christie_etal_19}. Therefore, we expect that the observed and synthetic light curves may show common features, such as the time-variability. 

\begin{figure}
  \centering
  \includegraphics[scale=0.31]{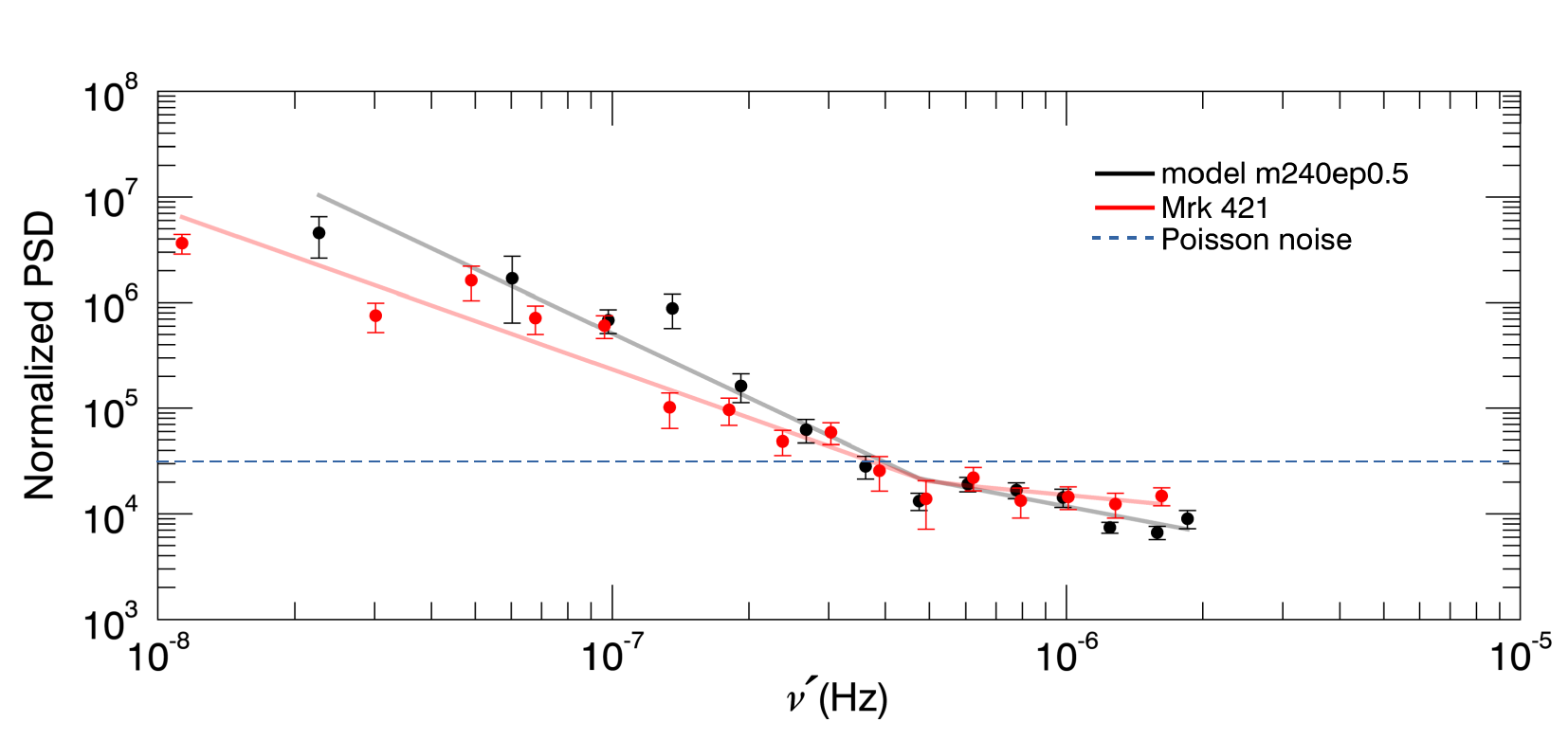}
  \caption{The diagram shows the normalized power spectral density (PSD) of the photon fluxes obtained from the light curve of Mrk~421 \citep[red line, obtained from][]{kushwaha_etal_17}, and our jet model m240ep0.5 (black line) in the observer frame. The blue traced line corresponds to the Poisson noise level for Mrk 421.}
  \label{fig:psd}
\end{figure}

We have then analyzed the PSD of the synthetic light curve and compared it with the one from Mrk~421 obtained by \citet{kushwaha_etal_17}. We obtained the synthetic PSD from the reference model m240ep0.5 during a time range between $t=39$ and $66$, where the kink instability achieves the nonlinear phase \citepalias[see also][]{medina_torrejon_etal2020}. Figure \ref{fig:psd} shows such comparison for the observed (red line) and synthetic (black line) PSD in the observer's frame. The latter shows a broken power-law profile\footnote{$f^\prime \propto \nu^{\prime a_{1}} (\nu^\prime < \nu^\prime_o)$ and $f^\prime \propto \nu^{\prime a_{2}} (\nu^\prime \ge \nu^\prime_{0}) $} with a frequency break $log_{10}(\nu^\prime_{0}) \sim -6.3$, in agreement with one obtained by \citet{kushwaha_etal_17}. We have also obtained a compatible power-law index ($a_2 \sim -0.6$) for frequencies higher than $\nu^\prime_0$\footnote{Nevertheless, we should point out that the  flattening observed for frequencies higher than $\nu^\prime_0$ is uncertain since the normalized PSD amplitudes are below the Poisson noise level (blue traced line in Figure \ref{fig:psd}).}. On the other hand, the synthetic PSD is steeper ($a_1 \sim -2.0$) than the observed one ($a_1 \sim -1.5$) for frequencies lower than $\nu_0^\prime$. Since the synthetic light curve is a result of turbulent magnetic reconnection events, the similarity found between the PSDs emphasizes the argument that the time-variability observed in Mrk~421 can be correlated with such events.

\section{Discussion and Conclusions}
\label{sec:conclusions}

In this work, we have evaluated how turbulent magnetic reconnection structures evolve in Poynting-flux dominated relativistic jets subject to the current-driven kink instability. To this aim, we have performed 3D-SRMHD simulations following the work of \citet{mizuno_etal_12} to obtain suitable initial and boundary conditions for a small portion of a comoving jet with an initial helical magnetic field and a moderate magnetization ($\sigma \sim 1.0$) near the jet spine. Motivated by the previous work of \citet{singh_etal_16}, we have examined the magnetic reconnection rates in such a system employing the algorithm developed in \citet{kadowaki_etal_18}, which was modified to take into account relativistic effects. 
This algorithm was able to perform a systematic search of all reconnection events in the system, allowing for  visualization of the local magnetic topology of the current sheets, as well as quantification of the reconnection rate and the magnetic power released in these events, in three different frames (the reconnection, the jet, and the observer's frame). Our results can be summarized as follows:

\begin{itemize}

\item The identification of the magnetic reconnection sites reveals that these are initially concentrated in the jet spine. At the beginning of the growth of the twists of the magnetic field driven the CDK instability (between $t \sim  20$ and $\sim 30~c.u.$), the current density structures are thick. During this phase (after $t \sim  25$), mostly sporadic, slow reconnection events with average rates smaller than $0.015 \tilde{V}_{A}$ are detected. As time evolves, after the instability grows exponentially up to saturation \citepalias[around $t=40$; see also][]{medina_torrejon_etal2020}, the jet wiggling structure amplitude continues to increase and becomes fully turbulent. The magnetic field lines are distorted and disrupted in several regions, and the associated current density structures become thinner (Figure \ref{fig:jetevol}), undergoing fast reconnection with the fastest rate around $0.23 \tilde{V}_{A}$ (at $t=56$). The system achieves a turbulent, quasi-steady-state regime between $t\sim 50$ and $66$ with an average rate of $0.051\pm0.026 \tilde{V}_{A}$ in agreement with the previous works of \citet{takamoto_etal_15} and \citet{singh_etal_16}. 

\item The  reconnection rate values resembles a log-normal distribution especially in the quasi-steady-state phase, after $t=50$ (Figure \ref{fig:vrec_hist}). Since log-normal profiles are signatures of turbulence, this is another indication that turbulence is driving the fast  reconnection events \citep[in agreement with the theory of][]{lazarian_vishiniac_99} along the relativistic jets. A $\chi^2$ analysis reveals that asymmetric reconnection events produce a deviation from the log-normal distribution, which indicates that such events maybe not be associated with the same process.

\item The convergence tests show no major differences in $\langle \tilde{V}_{rec} \rangle_s$ between the models with the middle and high resolutions ($240$ and $480$ cells in each direction), which implies that the fastest reconnection events are independent of the numerical resistivity, again in agreement with the turbulence-induced fast reconnection theory of \citet{lazarian_vishiniac_99} \citep[see also][]{kowal_etal_09,santos-lima_etal2020}, which predicts that the reconnection does not depend on the Ohmic resistivity of the environment.

\item Our qualitative analysis of $2D$ cuts of the local reconnection events (using the LIC method) evidences X-point-like structures of highly concentrated magnetic field lines (in a reconnection process) and magnetic islands (already reconnected) in several regions, where the magnetic field and the current density magnitudes are anti-correlated. These results are encouraging since they have demonstrated the efficiency of the algorithm at localizing events in different stages of reconnection. However, the $3D$ analysis has revealed that not always those magnetic islands are null point regions, where the magnetic field vanishes. Indeed, in the presence of the strong guide field (the helical magnetic structure), in some situations only one of the projected components is annihilated. This result indicates that analyses about the topology and evolution of blobs produced along relativistic jets due to reconnection events \citep[such as, e.g., in][]{giannios_etal_09,christie_etal_19} should be considered more carefully.

\item A quantitative analysis of the highest velocity values around each reconnection region (in the reconnection frame) indicates an Alfv\'{e}n speed value $\tilde{V}_{A} \sim 0.71c$, an inflow velocity $\tilde{V}_{in} \sim 0.075c$, and  an outflow velocity $\tilde{V}_{out} \sim 0.28c$. Furthermore, the reconnection (diffusion) region velocities are mildly relativistic with respect to the jet frame, where the Lorentz factor is not higher than $1.09$. Therefore, as we expected, a relativistic correction is appropriate for the Alfv\'{e}n speed, but it will not change significantly the velocity field in the jet frame. As we have seen, such correction becomes important only when we consider an observer's frame where the jet frame (or the computational domain) is moving relative to it with high bulk Lorentz factors ($\Gamma^\prime_j>10$).

\item The synthetic light curve built from the integrated magnetic reconnection power and evaluated from the reference model m240ep0.5 (assuming a typical Doppler factor $\delta^\prime \sim 5$) shows a high activity phase after $t \sim 39$ and coincides with the nonlinear phase of the growth of the CDK instability as well as the stage where several fast reconnection events are identified. The synthetic PSD extracted between $t=39$ and $66$ reveals a broken power-law profile with a frequency break $log_{10}(\nu^\prime_0)\sim -6.3$, in agreement with the observations of Mrk~421 (in the GeV band) obtained by \citet{kushwaha_etal_17}. These authors have also obtained a similar profile with PSD power-law indices $a_1 \sim -1.48 \pm 0.20$ (for $\nu^\prime \leq \nu^\prime_0$) and $a_2 \sim -0.57 \pm 0.48$ (for $\nu^\prime > \nu^\prime_0$). Though we have obtained for the lower frequency range    a  steeper power-law index for the synthetic PSD ($a_1\sim -2.0$), for the higher frequency range the value is very similar ($a_2\sim -0.6$). These results suggest that the turbulent fast magnetic reconnection driven by CDK instability is a possible process behind the daily time-variability observed in the GeV band in the blazar Mrk 421, as suggested by \citet{kushwaha_etal_17}. In future work, we have plans to apply similar study to observed TeV fast variability emission of other blazars (e.g., PKS 2155-304, Mrk 501, 3C 279, and 3C 54.3), in order to provide hints not only on the variability but also about the broken power law. The origin of this profile is still not  well understood in the observations and has been associated to the compact regions of the AGNs sources and also in galactic X-ray binaries \citep[see, e.g.,][]{mchardy_etal_05}. On the other hand, \citet{kushwaha_etal_17} have not found any evidence of a break scalability in the sources studied by them (NGC 1275, Mrk 421, B2 1520+31, and PKS 1510-089), which may indicate different origins. Thus, our numerical results can shed some light on this problem, strengthening reconnection models like e.g. the jets-in-a-jet model  \citep{giannios_etal_09, christie_etal_19}, or the  kink turbulence induced reconnection model here described \citepalias[and in][]{medina_torrejon_etal2020}, or even the stripped jet model \citep{giannios_uzdensky2019, zhang_giannios2021}.
\end{itemize}

The implications of our findings, specially for the VHE emission observed in Poynting-flux dominated relativistic jets are rather important. The reconnection search algorithm here employed can, in principle, be applied to any simulated system, including those with higher Lorentz factors, magnetization, and variability as, for instance, the observed blazars with TeV flares with time-scales of the order of  hundred seconds \citep[see, e.g.,][]{aharonian_etal_07, albert_etal_07b}, or GRBs \citep[e.g.,][]{drenkhahn_2002,giannios_spruit_2007,zhang_yan_11}. 
Employing this algorithm in a systematic study of 3D relativistic MHD jets subject to fast reconnection driven by, e.g., CDK as in this work, or other instabilities \citep[see e.g.,][]{nishikawa_etal_2020}, or by mechanisms like the mini jets-in-a-jet model \citep[e.g.][]{giannios_etal_09}, or the striped jet model \citep{giannios_uzdensky2019}, will allow to disentangle over these different processes and constrain their applicability. 
However, as stressed, the algorithm has limitations, even with the sample constraints applied in the present work specially in the identification of structures which are not reconnecting. More sophisticated methods can be used in future works. For instance, as discussed in Section \ref{sec:mag_structures}, the time-evolution analysis of the magnetic structure images is a way to identify reconnection sites at different stages. Despite the high number of images necessary to this method, machine learning techniques can be applied to recognize and automatically select such events. Recently, \citet{jafari_etal_2020} have presented a new method based on the correlation between the magnetic field-fluid slippage and the system's stochasticity level, that can be evaluated by the scalar field $\Psi ={1\over2}\boldsymbol{B}_l \cdot \boldsymbol{B}_L$, where $\boldsymbol{B}_l$ and $\boldsymbol{B}_L$ correspond to renormalized magnetic fields at different scales ($l$ and $L$). This  method also represents a powerful tool for the identification of reconnection structures in turbulent environments and future work should confront both techniques.

Finally, as stressed, this study has also provided the basis for us to explore, in a companion work, $in-situ$ particle acceleration in the reconnection sites identified along the simulated $3D$ relativistic jet \citepalias[see][]{medina_torrejon_etal2020}. We have found that thousands of protons injected with initial energies of 1 MeV are accelerated exponentially, predominantly parallel to the local magnetic field, achieving ultra-high energies, which could explain observed VHE and neutrino emissions in relativistic jets. This could be the case, for instance, of blazars like TXS~0506 +056 \citep{aartsen_etal_2018}, for which TeV gamma-rays and neutrinos have been simultaneously detected for the first time. In forthcoming work we plan to study in detail the radiative losses in these systems by incorporating radiative transfer effects to our model. 


\acknowledgments
The analyses and 3D SRMHD numerical simulations presented here were performed in the cluster of the Group of Plasmas and High-Energy Astrophysics (GAPAE), acquired with support from the Brazilian funding agency FAPESP (grant 2013/10559-5), and in the Blue Gene/Q supercomputer supported by the Center for Research Computing (Rice University) and Superintend\^{e}ncia de Tecnologia da Informa\c{c}\~{a}o da Universidade de S\~{a}o Paulo (USP). This work also made use of the computing facilities of the Laboratory of Astroinformatics (IAG/USP, NAT/Unicsul), whose purchase was also made possible by FAPESP (grant 2009/54006-4) and the INCT-A. LHSK also acknowledges support from  FAPESP (2016/12320-8),  EMdGDP  from FAPESP (grant 2013/10559-5) and the Brazilian agency CNPq (grant 308643/2017-8), and TEMT  from  CAPES. Y.M. is supported by the ERC Synergy Grant ``BlackHoleCam: Imaging the Event Horizon of Black Holes'' (Grant No. 610058). PK acknowledges ARIES Aryabhatta postdoctoral Fellowship (AO/A-PDF/770). Also, we would like to thank the anonymous referees for their useful comments and discussions.

\bibliography{bibliography.bib}



\end{document}

%% file: table00.tex
\begin{table*}
\centering
\caption{Parameters.}
\begin{tabular}{cccccccc}
\hline \hline

\multirow{2}{*}{Model} & & \multicolumn{2}{c}{Simulation} & & \multicolumn{3}{c}{Algorithm} \\ \cline{3-4} \cline{6-8}
 & & \multicolumn{1}{c}{Computational domain (c.u.)} & \multicolumn{1}{c}{Resolution (cells)} &  & \multicolumn{1}{c}{Subarray size (cells)} & \multicolumn{1}{c}{$\epsilon$} & \multicolumn{1}{c}{Edge position}

\\ \hline

m120ep0.5  & & $[6 , 6 , 6]$ & $[120,120,120]$ & & $[3,3,3]$ & $5.0$ & $0.5|\boldsymbol{J}_{max}| $ \\
m240ep0.5  & & $[6 , 6 , 6]$ & $[240,240,240]$ & & $[3,3,3]$ & $5.0$ & $0.5|\boldsymbol{J}_{max}| $ \\
m240ep0.1  & & $[6 , 6 , 6]$ & $[240,240,240]$ & & $[3,3,3]$ & $5.0$ & $0.1|\boldsymbol{J}_{max}| $ \\
m240ep0.05 & & $[6 , 6 , 6]$ & $[240,240,240]$ & & $[3,3,3]$ & $5.0$ & $0.05|\boldsymbol{J}_{max}|$ \\
m480ep0.5  & & $[6 , 6 , 6]$ & $[480,480,480]$ & & $[3,3,3]$ & $5.0$ & $0.5|\boldsymbol{J}_{max}| $

\\ \hline

\label{tab:parameters}
\end{tabular}
\end{table*}

%% file: table01.tex
\begin{table*}
\centering
\caption{Statistics of the $\Tilde{V}_{rec}$ distribution for the model $m240ep0.5$ ($t=50-66$).}
\begin{tabular}{lcccccc}
\hline \hline

\multirow{2}{*}{Statistics} & & \multicolumn{2}{c}{Symmetric and nonsymmetric profiles}            & & \multicolumn{2}{c}{Symmetric profiles}                           \\ \cline{3-4} \cline{6-7}
                            & & \multicolumn{1}{c}{Sample}  & \multicolumn{1}{c}{Fit (Lognormal)}  & & \multicolumn{1}{c}{Sample} & \multicolumn{1}{c}{Fit (Lognormal)} \\ \hline
Mean                        & & $0.047$                     & $0.05085 \pm 0.00060$                & & $0.048$                    & $0.04969 \pm 0.00038$               \\
$\sigma (\sqrt{\mathrm{Variance}})$  & & $0.024$                     & $0.02612 \pm 0.00082$                & & $0.022$                    & $0.02146 \pm 0.00049$               \\
Skewness                    & & $0.99$                      & $1.676   \pm 0.048$                  & & $1.26$                     & $1.376   \pm 0.029$                 \\
Kurtosis                    & & $1.99$                      & $5.38    \pm 0.33$                   & & $4.17$                     & $3.55    \pm 0.16$                  \\
$\chi^{2}_{red}$            & & $-$                         & $17$                                 & & $-$                        & $2.8$                               \\
Degrees of Freedom                      & & $-$                         & $78$                                 & & $-$                        & $76$                 

\\ \hline

\label{tab:fitvrec}
\end{tabular}
\end{table*}

%% file: table02.tex
\begin{table*}
\centering
\caption{Physical constant units (cgs) in the jet frame.}
\begin{tabular}{lcccccc}
\hline \hline

$\delta^{\prime}$ & $t_{0}$ (s) & $B_{0}$ (G) & $r_{0}$ (cm) & $\rho_{0}$ ($g/cm^3$) & $n_p$ ($\#/cm^3$) & $L_{0}$ ($erg/s$)  \\ \hline
$5.0$    & $1.3 \times 10^7$    & $3.7$                & $3.9 \times 10^{17}$  & $1.2\times 10^{-21}$           & $717$           & $4.8\times 10^{45}$ \\
$7.0$    & $1.8 \times 10^7$    & $1.3$                & $5.4 \times 10^{17}$  & $1.5\times 10^{-22}$           & $92$            & $1.2\times 10^{45}$ \\
$14$     & $3.6 \times 10^7$    & $0.16$               & $1.1 \times 10^{18}$  & $2.3\times 10^{-24}$           & $1.3$           & $7.2\times 10^{43}$ \\
$20$     & $5.2 \times 10^7$    & $0.06$               & $1.5 \times 10^{18}$  & $3.2\times 10^{-25}$           & $0.19$          & $2.1\times 10^{43}$ 

\\ \hline

\label{tab:punits}
\end{tabular}
\end{table*}